\documentclass[12pt,preprint]{aastex}

\newcommand{\sect}[1]{\S\,\ref{#1}}

\newcommand{\be}{\begin{displaymath}}
\newcommand{\ee}{\end{displaymath}}
\newcommand{\bea}{\begin{eqnarray}}
\newcommand{\eea}{\end{eqnarray}}

\newcommand\kms{\,km\,s$^{-1}$}

\shortauthors{Denissenkov, Pinsonneault, and MacGregor}
\shorttitle{Gravity Waves and Solar Uniform Rotation}

\begin{document}

\title{WHAT PREVENTS INTERNAL GRAVITY WAVES FROM DISTURBING THE SOLAR UNIFORM ROTATION?}

\author{Pavel A. Denissenkov\altaffilmark{1,2}, Marc Pinsonneault\altaffilmark{1}, and Keith B. MacGregor\altaffilmark{3}}
\altaffiltext{1}{Department of Astronomy, The Ohio State University, 4055 McPherson Laboratory,
       140 West 18th Avenue, Columbus, OH 43210; dpa@astronomy.ohio-state.edu, pinsono@astronomy.ohio-state.edu.}
\altaffiltext{2}{On leave from Sobolev Astronomical Institute of St. Petersburg State University,
   Universitetsky Pr. 28, Petrodvorets, 198504 St. Petersburg, Russia.}
\altaffiltext{3}{High Altitude Observatory, National Center for Atmospheric Research, P.O. Box 3000,
   Boulder, CO 80307-3000; kmac@hao.ucar.edu.}

\begin{abstract}
Internal gravity waves (IGWs) are naturally produced by convection in stellar envelopes, and they
could be an important mechanism for transporting angular momentum in the radiative interiors of stars.
Prior work has established that they could operate over a short enough time scale to explain the internal
solar rotation as a function of depth.  We demonstrate that the natural action of IGWs is to produce large scale
oscillations in the solar rotation as a function of depth, which is in marked contrast to the nearly uniform rotation
in the outer radiative envelope of the Sun.  An additional angular momentum transport mechanism is therefore
required, and neither molecular nor shear-induced turbulent viscosity is sufficient to smooth out the profile.
Magnetic processes, such as the Tayler-Spruit dynamo, could flatten the rotation profile.  We therefore conclude that
IGWs must operate in conjunction with magnetic angular momentum transport processes if they operate at all.
Furthermore, both classes of mechanisms must be inhibited to some degree by mean molecular weight gradients in
order to explain the recent evidence for a rapidly rotating embedded core in the Sun.
\end{abstract}


\section{Introduction}
\label{sec:intro}

During their pre-main sequence contraction, young solar-type stars are spun up
to rotational velocities of the order of 100\,\kms. However, during their subsequent
main sequence (MS) evolution the surface rotation slows down as a result of angular momentum loss
through magnetized stellar winds (\citealt{k88,mp08}). If a convective envelope and
a radiative core of a solar-type MS star rotated independently of one another then the surface spindown
would lead to a strong differential rotation beneath the envelope. In contradiction with
this, helioseismic data (e.g., \citealt{cea03}) reveal that the solar radiative core
rotates as a solid body and almost synchronously with the convective envelope
at least down to the radius $r\approx 0.2\,R_\odot$.
Besides, the spindown of young cluster stars (\citealt{sh87,kea95,bea97,kea97,b03})
requires that the internal differential rotation only persists for
timescales of the order of 20 Myr to 100 Myr. These data indicate that, in radiative interiors of solar-type MS stars,
there is an efficient mechanism of angular momentum redistribution that couples their core and
envelope rotation. Unfortunately, its physical nature remains elusive in spite of many
attempts to understand it made in the last years. Matters are further complicated by the requirement that
an appropriate model of angular momentum transport in the solar-type MS stars should also reproduce
the intricate variations (depletion) of the surface Li abundance as functions of age and effective temperature
observed in the same stars (\citealt{sr05}).

A breakthrough in solving this complex problem has recently been announced by \cite{tch05} and
\cite{cht05}. They have shown how internal gravity waves (hereafter, IGWs) generated by
the envelope convection can extract angular momentum from rapidly rotating radiative cores of
the solar-type MS stars on short enough timescales to explain both the spindown of young cluster stars
and the quasi-solid-body rotation of the Sun. Furthermore, they have found that the quick 
flattening of the internal rotation profile
by IGWs reduces element diffusion coefficients associated with hydrodynamic instabilities induced by
differential rotation to values consistent with those constrained by the Li data.

Pursuing the goal of uncovering intrinsic causes of canonical extra mixing in low-mass red giant branch stars
(\citealt{dvb03}) and trying to understand the origin of fast rotation of red horizontal branch stars (\citealt{sp00}),
we have been undertaking a critical review of different transport mechanisms in
stellar radiative zones. When applying our test model of angular momentum redistribution by IGWs to a model of the present-day Sun,
we have found that IGWs strongly disturb the solar internal rotation making it disagree with the helioseismic data.
The disturbance can be eliminated only if there is another transport mechanism competing with IGWs. Its efficiency should
exceed that of the rotational shear mixing used by \cite{tch05} by more than three orders of magnitude. Alternatively,
a magnetic transport mechanism could be a competitor for IGWs. But in that case the mechanism preventing
IGWs from disturbing the solid-body rotation of the Sun could itself be responsible for both shaping the
internal rotation of the solar-type MS stars and assisting in the depletion of their surface Li abundance. This paper presents
a discussion of computational results that support our conclusions.

\section{Angular Momentum Transport by IGWs}
\label{sec:transport}

In a gravitational field, any perturbation exerted on a fluid excites in it both acoustic (p-modes) and
internal gravity waves (g-modes). The driving force for the latter is the buoyancy, as opposed to the pressure for the p-modes.
A general discussion of their combined linear theory can be found in the book by \cite{l78}.
Our application of the IGW theory first addresses the following three questions: How do IGWs propagate through radiative layers?
How do they redistribute angular momentum in a rotating star? How are IGWs generated? Answering the last question necessarily includes
a discussion of the IGW spectrum. We start general comments on IGWs, then attempt to examine the underlying issues.               

The physics of angular momentum transport and chemical mixing by IGWs in stellar radiative zones 
has been comprehensively described by \cite{p81}, \cite{gls91}, \cite{zea97}, \cite{r98}, \cite{kea99}, \cite{ms00}, \cite{kmg01},
\cite{tea02}, and \cite{tch05}. In all of these papers, IGWs are considered to be generated by large-scale 
turbulent fluid motions either in the convective envelope or at the interface between
the radiative core and convective envelope. The net energy flux of IGWs at the core/envelope interface $r_{\rm c}$ is usually estimated
as $F_E(r_{\rm c})\approx {\cal M}_{\rm t}F_{\rm c}$, where $F_{\rm c}=\rho_{\rm c}v_{\rm c}^3\approx 0.1\,(L/4\pi r_{\rm c}^2)$ is
the convective flux, $L$ is the star's luminosity, and ${\cal M}_{\rm t}=\omega_{\rm c}/N_{\rm c}$ is the turbulent Mach number.
In the last ratio, $N_{\rm c}$ is the buoyancy frequency immediately beneath
the interface, while $\omega_{\rm c}=2\pi v_{\rm c}/\lambda_{\rm c}$ is the turnover frequency 
of the largest convective eddy approaching the interface with the velocity $v_{\rm c}$. 
Everywhere in this paper $\omega$ denotes the circular frequency. Although it should be measured in rad\,s$^{-1}$, we will 
always express its values in units of $\mu$Hz, assuming that $1\,\mu$Hz\,$\equiv 10^{-6}$\,rad\,s$^{-1}$. 
In the mixing-length theory (MLT) of convection, that we apply,
the diameter of the largest turbulent eddy, which also measures its mean free path $\lambda_{\rm c}$, is 
an $\alpha_{\rm MLT}$ fraction of the local pressure scale height $H_P$. 
We employ the stellar evolution code described by \cite{dea06}. Gravitational settling is not included because it does not affect
the IGW propagation. Indeed, IGWs can easily penetrate inner radiative cores of low-mass MS stars (\citealt{tch05}) where radial variations of the mean molecular weight $\mu$
are much stronger than those incurred from the operation of gravitational settling. We use the \cite{gn93} mixture of
heavy elements. Our calibrated solar model reproduces the solar luminosity
($L_\odot = 3.85\times 10^{33}$\,erg\,s$^{-1}$) and radius ($R_\odot = 6.96\times 10^{10}$\,cm) at
the solar age of 4.57 Gyr; this procedure yields the helium and heavy-element mass fractions
$Y = 0.273$, $Z = 0.018$, and the mixing length $\alpha_{\rm MLT}$ of 1.75.
Our solar model has $r_{\rm c}\approx 0.713\,R_\odot$ and ${\cal M}_{\rm t,\odot}\approx 8.3\times 10^{-4}$, so that the energy luminosity of
IGWs at the bottom of its convective envelope $L_E(r_{\rm c})=4\pi r_{\rm c}^2F_E(r_{\rm c})$ comprises only 
$\sim$\,0.0083\,\% of $L_\odot$.

On their way from the bottom of the convective envelope toward the center of a solar-type MS star,
IGWs experience radiative damping. This can be taken into account by applying a wave attenuation factor
$\exp(-\tau)$ to $L_E(r_{\rm c})$. To calculate the effective optical depth, we use the relation
\bea
\tau = [l(l+1)]^{3/2}\int_r^{r_{\rm c}}K\frac{NN_T^2}{\sigma^4\sqrt{1-(\sigma/N)^2}}\frac{dr'}{r'^3}
\label{eq:tau}
\eea 
derived by \cite{zea97}. Here, 
$N^2 = N_T^2 + N_\mu^2$,
where
\be
N_T^2 = \frac{g\delta}{H_P}(\nabla_{{\rm ad}}-\nabla_{{\rm rad}}),
\ \ \mbox{and}\ \
N_\mu^2 = g\varphi\,\left|\frac{\partial\ln\mu}{\partial r}\right|
\ee
are the $T$- and $\mu$-component of the square of the
buoyancy frequency. In the last two expressions,
$\nabla_{\rm rad}$ and $\nabla_{\rm ad}$ are the radiative
and adiabatic temperature gradients (logarithmic and with respect to
pressure), and $g$ is the local gravity.
The quantities
$\delta =
-\left(\partial\ln\rho/\partial\ln T\right)_{P,\,\mu}$ and
$\varphi = \left(\partial\ln\rho/\partial\ln\mu\right)_{P,T}$
are determined by the equation of state. In our IGW computations,
the equation of state for the ideal gas is used. In this particular case, $\delta = \varphi = 1$.
Also in equation (\ref{eq:tau}), $K = 4acT^3/3\kappa\rho^2C_P$
is the radiative diffusivity with $\kappa$ and $C_P$ representing the Rosseland mean opacity and
the specific heat at constant pressure, respectively. The quantities $l$ and $\sigma$ are introduced below.

Turbulent eddies with different length and overturn time scales present in the convective envelope 
can generate a whole spectrum $S_E(r_{\rm c},l,m,\omega)$ of IGWs
with different spherical degrees $l\geq 1$, azimuthal numbers $m$ ($|m|\leq l$), and frequencies $\omega$.
If the star rotates
and its angular velocity $\Omega$
varies with $r$ then the optical depth (\ref{eq:tau}) depends on all three of the wave's
spectral characteristics,
the latter two entering it through the doppler-shifted frequency
$\sigma = \omega - m[\Omega(r)-\Omega(r_{\rm c})]$ 
(we use a cartesian coordinate system in which
the z-axis is colinear with the vector $\mathbf{\Omega}$).
Values of this frequency should
be watched to remain between 0 and $N$. When $\sigma\rightarrow 0$, the optical depth $\tau$
approaches the infinity, and the wave is completely absorbed by the surrounding medium.
On the other hand, when $\sigma\rightarrow N$, the wave is totally reflected back (\citealt{r98}).
If only the IGWs with $1\leq l\leq l_{\rm max}$ and $\omega_{\rm min}\leq\omega\leq\omega_{\rm max}$
are excited and propagate into the radiative core then their net energy flux at the core/envelope interface is
\bea
F_E(r_{\rm c}) = \sum_{l=1}^{l_{\rm max}}\sum_{m=-l}^{l}\int_{\omega_{\rm min}}^{\omega_{\rm max}}S_E(r_{\rm c},l,m,\omega)d\omega = {\cal M}_{\rm t}F_{\rm c}.
\label{eq:neteflux}
\eea

Besides energy, IGWs can also carry angular momentum the flux of which is
\bea
F_J = \frac{m}{\sigma}F_E
\label{eq:fjfe}
\eea
(\citealt{r98,kea99}). 
Since both $F_E$ and $\sigma$ are positive, this relation means that
prograde waves (those with $m > 0$) carry positive angular momentum while
retrograde waves ($m < 0$) transport negative momentum.
Note that before Ringot's elucidating paper the improper (negative) sign was used in equation (\ref{eq:fjfe})
by many researchers creating some confusion in the field.

Let us choose a frame of reference co-rotating with
the convective envelope and denote $\Delta\Omega(r)\equiv\Omega(r)-\Omega(r_{\rm c})$. 
We will assume that, in spite of the action of the
Coriolis force in this noninertial frame of reference, the spectrum of IGWs generated by
the envelope convection is still axisymmetric, i.e. $S_E(r_{\rm c},l,m,\omega) = S_E(r_{\rm c},l,-m,\omega)$;
if necessary, corrections due to the Coriolis force can be taken into account later on (e.g., \citealt{tch03}).
If $\Delta\Omega(r)=0$, i.e. if the whole star rotates uniformly, then $\sigma = \omega$, and $\tau$ does not depend on $m$.
In this case, the total angular momentum luminosity associated with IGWs
\bea
L_J(r) = 4\pi r_{\rm c}^2\,\sum_{l=1}^{l_{\rm max}}\sum_{m=-l}^{l}m\int_{\omega_{\rm min}}^{\omega_{\rm max}}
S_E(r_{\rm c},l,m,\omega)\exp\{-\tau(r,l,m,\omega)\}\,\frac{d\omega}{\omega}
\label{eq:netjflux}
\eea
is equal to zero at any radius $r$ below the convective envelope.

On the other hand, if $\Delta\Omega(r)\not= 0$, i.e. if the star rotates differentially,
then IGWs will experience selective damping in the radiative core. Indeed, let us assume
for example that $\Delta\Omega(r)$ increases with depth (when $r$ decreases).
In this case, the frequency $\sigma$ of a prograde wave with spectral characteristics $l$, $m$, and $\omega$ will be doppler shifted to values
smaller than the intrinsic frequency $\omega$, while that of a retrograde wave with the characteristics $l$, $-m$ and $\omega$ will
become larger than $\omega$ as the waves propagate inward. Hence, the difference
between their attenuation factors $\exp\{-\tau(r,l,m,\omega)\} - \exp\{-\tau(r,l,-m,\omega)\}$
will diminish with the depth. This happens because the prograde waves carrying the positive angular momentum
experience stronger damping (if $\Delta\Omega(r) > 0$) on their way into the radiative core 
than the retrograde waves carrying the negative angular momentum.
The positive angular momentum absorbed by the surroundings at the beginning of the waves' path will spin up
rotation locally compared to $\Omega(r_{\rm c})+\Delta\Omega$ while the negative momentum accumulating in the waves
as they advance inward will be deposited deeper where the retrograde waves get eventually absorbed.
This briefly sketched physics of angular momentum redistribution by IGWs is incorporated into
the following PDE:            
\bea
\rho r^2\,\frac{\partial\,\Delta\Omega}{\partial t} =
\frac{1}{r^2}\frac{\partial}{\partial r}\left(\rho r^4\nu\,\frac{\partial\,\Delta\Omega}{\partial r}\right) +
\frac{3}{8\pi}\,\frac{1}{r^2}\frac{\partial L_J}{\partial r},
\label{eq:ampdf}
\eea
where $\nu$ is a viscosity.
Supplemented with eqs. (\ref{eq:tau}), (\ref{eq:netjflux}), an expression for
the IGW spectrum at $r=r_{\rm c}$, and appropriate initial and boundary conditions,
this equation describes how the star's internal rotation profile evolves
in the presence of IGWs and a viscous force.

\section{Spectra of IGWs}
\label{sec:spectrum}

Given that the optical depth $\tau$, that determines the efficiency of damping of an IGW, strongly depends
on the wave's spectral characteristics $l$, $m$, and $\omega$, knowing 
the spectral energy distribution for IGWs at the core/envelope interface
is as important as estimating their net energy. In this paper, we will use two analytical prescriptions 
(eqs. \ref{eq:zea97} and \ref{eq:tea02} below) for
the spectra of IGWs. They correspond to two different physical mechanisms of the IGW excitation by turbulence
in the convective envelope. The formal lower and upper limits for the intrinsic frequency of these IGWs
are $\omega_{\rm min} = \omega_{\rm c}$ and $\omega_{\rm max} = N_{\rm c}$. 

Unfortunately, multidimensional hydrodynamic simulations of turbulent convection in
the solar envelope and its penetration into the radiative core give contradictory results
on the spectrum of IGWs generated in these numerical experiments.
For example, in their 2D simulations \cite{rg05} have found an IGW energy flux evenly distributed in
frequency at least for $l\leq 20$ with a peak energy that is three orders of magnitude smaller than
that predicted by equation (\ref{eq:tea02}). On the other hand, \cite{kea05} claim that
the broad frequency IGW spectrum reported in their earlier
publication was an artefact of the 2D approximation. They emphasize that the IGW flux obtained in
their new 3D simulations is of the same order as the one calculated from the simple parametric model of
\cite{gls91} based on the MLT. This finding encourages us to use spectrum
(\ref{eq:zea97}) as our primary IGW model.

\subsection{The Garc\'{\i}a L\'{o}pez \& Spruit Spectrum}

This prescription approximates the spectrum of IGWs generated at the core/envelope interface
as a result of dynamic hitting at the radiative side of the interface by breaking convective eddies.
This approximation was proposed and developed by \cite{p81}, \cite{gls91}, and \cite{zea97}. We use an expression derived in the last work
\bea
S_E(r_{\rm c},l,m,\omega) = \frac{1}{2}\,\rho_{\rm c}v_{\rm c}^3\,\frac{\omega_{\rm c}}{N_{\rm c}}\,\frac{1}{l_{\rm c}\,\omega_{\rm c}}
\left(\frac{\omega_{\rm c}}{\omega}\right)^3\frac{1}{2l}.
\label{eq:zea97}
\eea
Equation (\ref{eq:zea97}) is obtained under the following assumptions: {\it (i)} the dynamic pressure in the waves of a frequency $\omega$ 
matches that produced by the convective eddies with the overturn time $\sim$\,$\omega^{-1}$;
{\it (ii)} the kinetic energy spectrum of the convective motions in the envelope is represented
by the Kolmogorov law; {\it (iii)} besides waves at their own length scale $\lambda$, the convective eddies
also excite IGWs with horizontal wave lengths $\lambda_{\rm h} > \lambda$ by the superposition of their incoherent
action on the interface; these longer waves have a velocity amplitude reduced by the factor $\lambda/\lambda_{\rm h}$ (\citealt{gls91}).
In this prescription, $l_{\rm max} = l_{\rm c}(\omega/\omega_{\rm c})^{3/2}$, where $l_{\rm c} = 2\pi r_{\rm c}/\lambda_{\rm c}$. 
Applying the integration and summation (\ref{eq:neteflux}) to the spectrum (\ref{eq:zea97}) and noticing
that $\omega_{\rm c}\ll N_{\rm c}$, we find that $F_{\rm E}(r_{\rm c}) = {\cal M}_{\rm t}F_{\rm c}$ as expected.

\subsection{The Goldreich et al. Spectrum}

The second prescription originated from the investigation of the stochastic excitation of p-modes
by turbulent convection in the Sun carried out by \cite{gea94}. In this model, IGWs are
generated by the fluctuating Reynolds stresses produced by turbulent fluid motions in the convective envelope.
Following \cite{tea02}, the emerging IGW spectrum at the core/envelope interface can be estimated as
\bea
S_E(r_{\rm c},l,m,\omega) & = & \frac{1}{2l}\,\frac{\omega^2}{4\pi}\int_{r_{\rm c}}^{R_{\rm c}}dr\,\frac{\rho^2}{r^2}
\left[\left(\frac{\partial\xi_r}{\partial r}\right)^2 + l(l+1)\left(\frac{\partial\xi_{\rm h}}{\partial r}\right)^2\right] \nonumber \\
& & \times\exp\left[-l(l+1)\frac{h_\omega^2}{2r^2}\right]
\frac{v^3\lambda^4}{1+(\omega\tau_\lambda)^{15/2}},
\label{eq:tea02}
\eea
where $R_{\rm c}$ is the outer radius of convective envelope, $\tau_\lambda = \lambda/v$ is the overturn time of
convective elements of the size $\lambda = \alpha_{\rm MLT}H_P$ moving with the velocity $v$ (at $r=r_{\rm c}$,
$\lambda$ and $v$ coincide with the parameters $\lambda_{\rm c}$ and $v_{\rm c}$ defined in \sect{sec:transport}),
and $h_\omega = \lambda\min\{1,\,(2\omega\tau_\lambda)^{-3/2}\}$. Like equation (\ref{eq:zea97}),
the latter equation has been derived under the assumption that the turbulent motions in the convective envelope
obey the Kolmogorov law. 

The radial displacement wave function $\xi_r$ and the horizontal one $l(l+1)\xi_{\rm h}$, 
the latter being related to the former by the continuity equation (\citealt{zea97}), are normalized
to the unit IGW energy flux just below the convection zone.
For the radial function, we use the WKB solution (\citealt{p81,zea97,kea99})
\bea
\xi_r(r) = C[l(l+1)]^{1/4}\frac{\exp\left[-\int_{r_{\rm c}}^r\frac{|N|}{\omega}
\frac{\sqrt{l(l+1)}}{r'}dr'\right]}{\omega(\rho\, r\, |N|\,)^{1/2}}.
\label{eq:ksir}
\eea
This equation shows that in the convective envelope, where $N^2<0$, IGWs are evanescent. Therefore only those of them
that are excited close to the core/envelope interface will effectively contribute to the IGW flux at $r=r_{\rm c}$. 
The constant $C$ is adjusted in our computations
for the spectrum (\ref{eq:tea02}) to yield $F_E(r_{\rm c}) = {\cal M}_{\rm t}F_{\rm c}$ with $F_{\rm c} = 0.1\,(L/4\pi r_{\rm c}^2)$.
The factor $(2l)^{-1}$ in eqs. (\ref{eq:zea97}) and (\ref{eq:tea02})
takes into account the assumed energy equipartition between the wave counterparts with opposite signs of the azimuthal number ($-l\leq m\leq l$).

\begin{figure}
\plotone{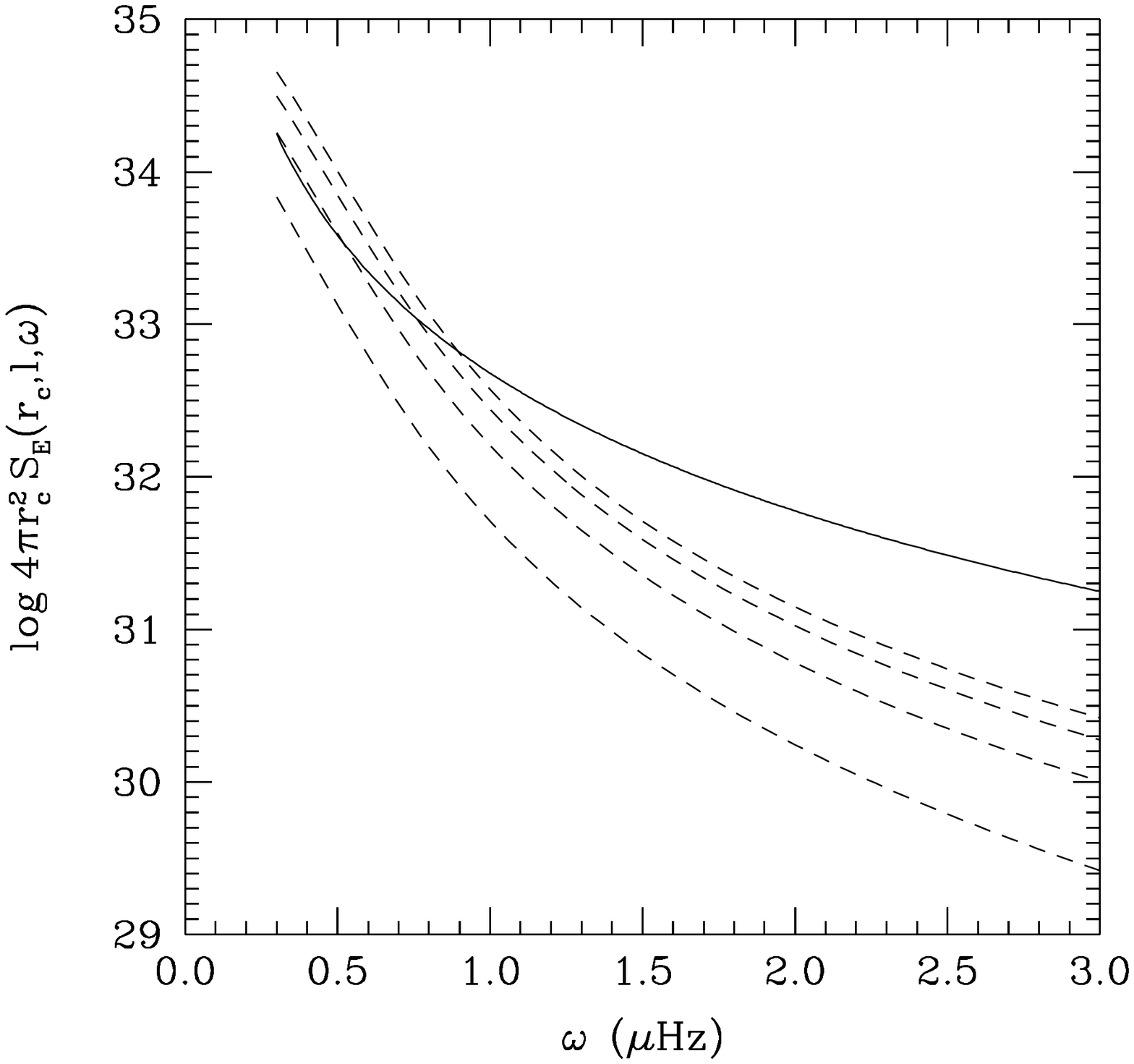}
\caption{Logarithms of the energy luminosity of IGWs at the Sun's
         core/envelope interface summed over all available azimuthal 
         numbers ($m=-l,\ldots,l$) as functions of $l$ and $\omega$.
         Solid curve corresponds to spectrum (\ref{eq:zea97}),
         dashed curves --- to spectrum (\ref{eq:tea02}) for $l=1,\,2,\,3$,
         and 4 (from the lower to upper curve).
        }
\label{fig:f1}
\end{figure}

In Fig.~\ref{fig:f1}, we have plotted the logarithms of the IGW energy luminosity summed
over all available values of $m$ at the core/envelope interface
in our solar model. Those were calculated using the spectra (\ref{eq:zea97}) (solid curve) and (\ref{eq:tea02}) 
(dashed curves) with the same value of ${\cal M}_{\rm t}=8.3\times 10^{-4}$ that gives $L_E(r_{\rm c})\approx 3.2\times 10^{29}$\,erg\,s$^{-1}$. 
Apparently, they have quite different
dependences on the spherical degree and frequency. Whereas the first spectrum
does not depend on $l$ at all\footnote{The factor $(2l)^{-1}$ disappears after
the summation over all $m=-l,\ldots,l$.} 
(eq. \ref{eq:zea97}), the second is estimated to be proportional to $l^p$, where
$p\approx 1.6$ for the first 4 degrees. 
Besides, the first spectrum declines with increasing $\omega$ much slower (with a power $-3$) than the second
spectrum (with a power $\sim$\,$-4.5$).   
As a result, in the second case the dipole wave ($l=1$), that experiences the least damping (eq. \ref{eq:tau}),
carries much less energy toward the radiative core than it does in the first case, 
especially at higher frequencies (compare the lower dashed and solid curve).

\subsection{Uncertainties in the MLT}

\begin{figure}
\plotone{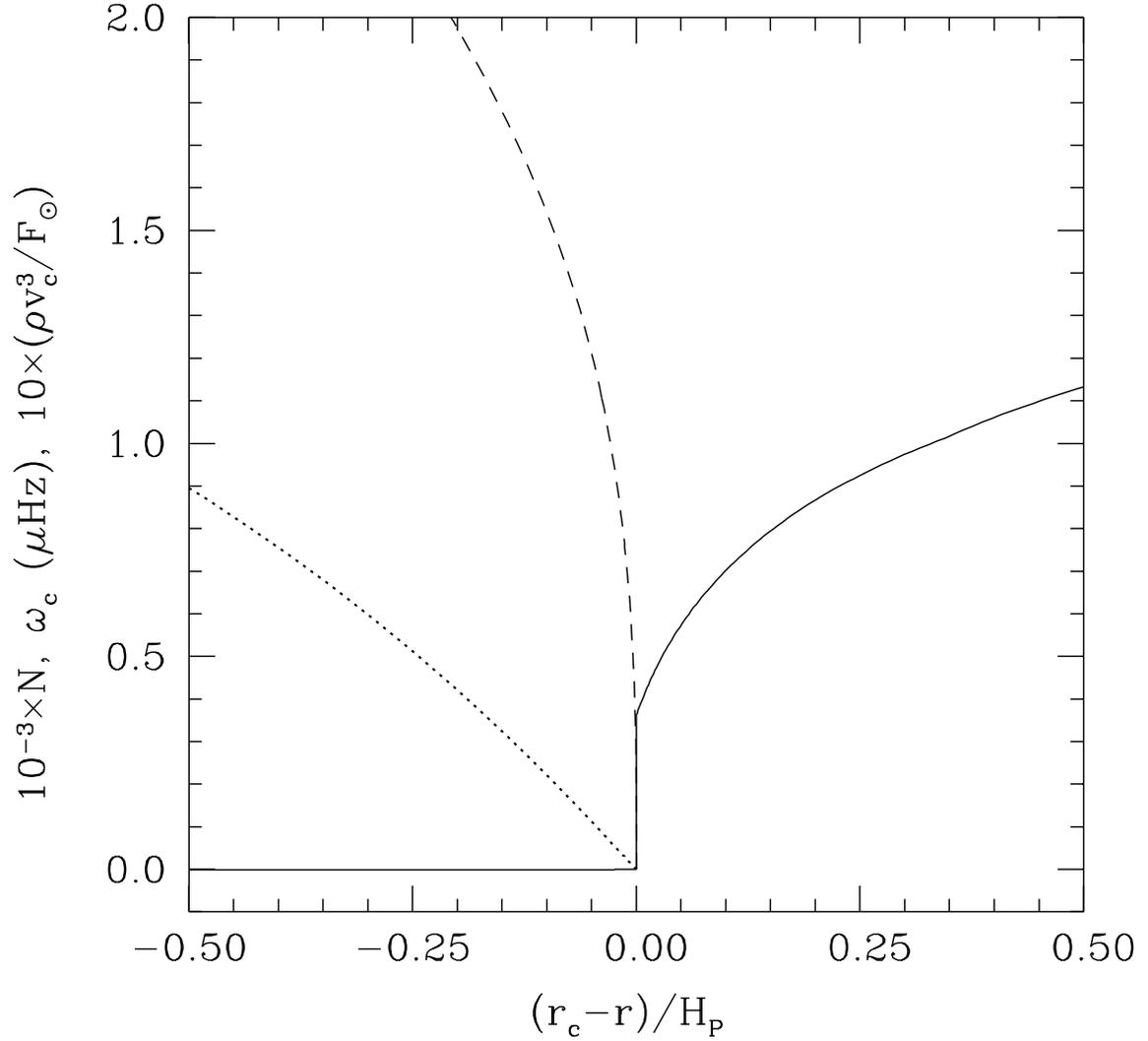}
\caption{The convective overturn frequency $\omega_{\rm c}$ (dashed curve),
         the buoyancy frequency $N$ (solid curve), and
         the ratio of the convective flux $\rho v_{\rm c}^3$
         to the total flux $F_\odot$ (dotted curve) as functions of the relative depth
         (expressed in units of the pressure scale height) on the both sides of
         the Sun's core/envelope interface.
        }
\label{fig:f2}
\end{figure}

Fig.~\ref{fig:f2} shows how the convective and buoyancy frequency, $\omega_{\rm c}$ and $N$, vary 
with the radius on the opposite sides of the core/envelope interface in our solar model. 
From this figure, it is evident that neither the minimum frequency of IGWs $\omega_{\rm min}=\omega_{\rm c}$ nor
the ratio ${\cal M}_{\rm t}=\omega_{\rm c}/N_{\rm c}$ estimating the net energy flux 
$F_E(r_{\rm c})\approx{\cal M}_{\rm t}F_\odot$
(dotted curve in the figure shows that the latter approximation becomes valid at $r\ga r_{\rm c}+0.5H_P$),
can be predicted with a confidence by the MLT of convection employed by us. 
The uncertainties are caused by the rapid growth of $\omega_{\rm c}$
and $N$ with an increasing distance from the interface. Strictly speaking, both $\omega_{\rm c}\propto v_{\rm c}$
and $N\propto |\nabla_{\rm rad} -\nabla_{\rm ad}|^{1/2}$ should vanish at $r=r_{\rm c}$.
However, as soon as we step aside from the interface, both of them jump up to finite values.
In stellar model computations, their first nonzero values depend on spatial resolution:
the higher the resolution is, the smaller these values are. For example,
in our computations of the present-day Sun's model, the results of which are plotted in Fig.~\ref{fig:f2},
we find $\omega_{\rm c}\approx 0.30\,\mu$Hz and $N_{\rm c}\approx 360\,\mu$Hz, hence ${\cal M}_{\rm t}\approx 8.3\times 10^{-4}$.
If we had taken into account convective overshooting beyond the formal lower boundary of convective envelope
located at the radius $r=r_{\rm c}$, where $\nabla_{\rm rad}=\nabla_{\rm ad}$ (the Schwartzschild criterion),
then we would have obtained a larger value of $N_{\rm c}$. In a value of ${\cal M}_{\rm t}=\omega_{\rm c}/N_{\rm c}$,
the increase of $N_{\rm c}$ due to the overshooting can partly be compensated
by choosing a larger representative value of $\omega_{\rm c}$ that should account of its rapid growth with radius
on a length scale much less than $H_P$ immediately above the core/envelope interface (dashed curve).

Besides the aforementioned uncertainties in the choice of representative values for $\omega_{\rm c}$ and $N_{\rm c}$,
the MLT does not account for the fact, established by laboratory experiments, observations in
the Earth's atmosphere and numerical simulations (e.g., see \citealt{sn89,cea91,rz95,ms00,rg06}), that
downward flows in a strongly stratified convection zone are much more energetic and confined in their
horizontal extent than upward flows. This may result in an underestimate of the IGW net energy flux, especially
in stars with deep convective envelope, like the Sun (\citealt{kea05}).

Given the uncertainties in the MLT and the fact that we want to investigate the           
stability of the solar uniform rotation against perturbations by the IGWs that were supposed to be powerful enough
to produce that uniform rotation in the past, we tentatively choose $\omega_{\rm min}=\omega_{\rm c}=0.30\,\mu$Hz 
while considering the turbulent Mach number ${\cal M}_{\rm t}$
as a free parameter with its minimum value equal to our estimated ratio 
$\omega_{\rm c}/N_{\rm c}\approx 8.3\times 10^{-4}$. This may actually put a conservative lower limit on the energetics of such IGWs
because other authors, including \cite{cht05} who demonstrated how the IGWs could shape the Sun's uniform rotation,
employed larger values of ${\cal M}_{\rm t}$ and sometimes also a higher $\omega_{\rm min}$.
Therefore, their low-degree high-frequency waves, that can penetrate deep into the radiative core,
carried more kinetic energy than they do in our basic case,
assuming, of course, that we use the same IGW spectrum. For the convective parameter 
$l_{\rm c}$, we use an MLT value $l_{\rm c}=30$ from our calibrated solar model.

\section{Qualitative Description of Expected Solutions}

After the proper (positive instead of negative) sign in relation (\ref{eq:fjfe}) had been defined
by \cite{r98} it became clear that the addition of IGWs to other processes
responsible for the redistribution of angular momentum in stellar radiative zones would not be a trivial problem.
Indeed, instead of resulting in an exponential decay of perturbations of the internal rotation profile
toward a solid body rotation (e.g., \citealt{kq97}), the damping of IGWs leads to progressively growing deviations of
local rotation from its initial state, even if this state is close to the solid body rotation. 
This is caused by the fact that in a region rotating faster
than the convective envelope the prograde waves experience stronger damping than the retrograde waves,
which results in a deposit of positive angular momentum there and hence in a further spin-up of this region.
The opposite is true for a region rotating slower than the convective envelope. Of course, the perturbations of
the rotation profile cannot grow infinitely large. First of all, an excess of (positive or negative) angular momentum in them
can slowly be dissipated through the molecular viscosity. Second and more important, when the rotational shear produced by the
perturbations becomes strong enough to trigger a shear instability
the turbulent viscous friction associated with the shear-induced turbulence will start to 
contribute to smoothing the perturbations out.

Apparently, the outcome of the competition between the disturbing action of IGWs and the smoothing effect of viscous friction
depends on their relative strength. Let us take a constant viscosity $\nu =\nu_0$. It is obvious that,
as long as $\nu_0$ is kept extremely large, the rotation profile will remain stationary because
any perturbation of $\Omega$ by IGWs will be quickly neutralized by the viscous dissipation.
Of course, this does not preclude gradual temporal changes of the rotation profile as a whole
due to the redistribution of angular momentum by IGWs as described by eq. (\ref{eq:ampdf}).

If we begin to decrease $\nu_0$ then at some critical value of it, which is proportional to
the local IGW energy flux, regular oscillations of $\Omega$ will set in.
In the Earth's atmosphere, this bifurcation of a stationary solution toward an oscillatory solution
is believed to be observed experimentally as the quasi-biennial oscillations (QBO) of the mean zonal
wind in the equatorial stratosphere disturbed by IGWs coming down from the troposphere
(e.g., \citealt{lh68,p77,yh88}). In models of the solar-type MS stars, a behavior of the $\Omega$-profile resembling that of the QBO
has been found as a solution of equation (\ref{eq:ampdf}) near the top of radiative core by 
\cite{kea99,kmg01,tea02,tch03}, and \cite{cht05}. \cite{tch05} have
called it the shear layer oscillations (SLOs). They have proposed that the SLOs
work as a filter for IGWs on their way toward the radiative core. If $\Omega$ increases with the depth in the core, as is expected
in a solar-type MS star losing its angular momentum from the surface through a magnetized stellar wind,
then the SLOs predominantly filter out the retrograde waves. These are absorbed closer to the center.
Possessing the minimum angular momentum, it is the very central part of the core that slows down the first.
Following it, increasingly more and more distant from the center layers get successively decelerated (\citealt{tch05}).
Note, however, that this theoretical prediction has recently been challenged by results of
a preliminary analysis of the GOLF data on solar g-mode oscillations reported by \cite{gea07}.
They suggest that the solar inner core at $r\la 0.15\,R_\odot$ may rotate three to five times as fast as
the rest of the radiative zone. If these results are confirmed by a further analysis they will likely rule out
the angular momentum redistribution in the Sun by IGWs.

If we continue to reduce $\nu_0$ further on then the oscillations will grow up in
amplitude until finally they will turn into chaotic variations of $\Omega$ (e.g., \citealt{kmg01}).

\section{Results: Two Spectra, Five Viscosities}
\label{sec:results}

\begin{figure}
\plotone{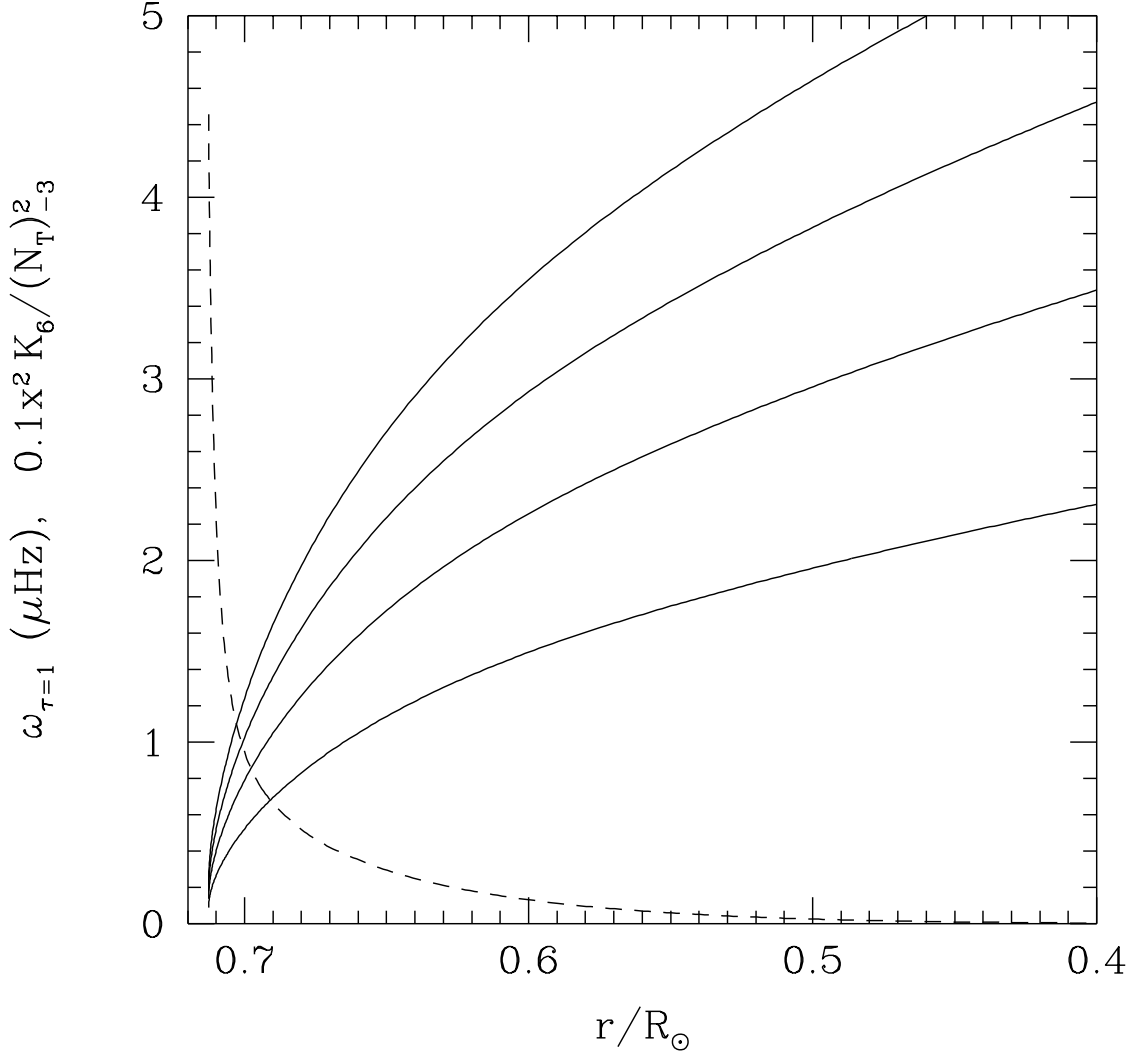}
\caption{The frequency $\omega_{\tau=1}$ for which
         the effective optical depth (eq. \ref{eq:tau}) in the IGW attenuation 
         factor $\exp(-\tau)$ equals to one at a given radius in the case of
         uniform rotation (from the lower to upper solid curve:
         $l$ increases from 1 to 4, and $\omega_{\tau=1}\propto [l(l+1)]^{3/8}$).
         Dashed curve is the normalized stellar structure parameter in expression
         (\ref{eq:nuv0}) for the shear-induced viscosity.
        }
\label{fig:f3}
\end{figure}

In this section, we present and discuss results of our numerical solutions of
the PDE (\ref{eq:ampdf}) that have been obtained with the IGW spectra
(\ref{eq:zea97}) and (\ref{eq:tea02}) using the viscosity prescriptions
summarized in Appendix~B. Besides restricting the azimuthal number
$m$ by the values of $\pm l$, we have also considered  a limited set of
the spherical degree. In most of our computations, we have only used a set of numbers
$l=1,\,2,\,3$, and 4. We do not think that adding higher degrees 
would qualitatively change our results and conclusions. 
Indeed, the optical depth in the wave attenuation factor $\exp(-\tau)$
increases as $\tau\propto l^3$ for $l > 1$, therefore at a same radius in the Sun's radiative core
a contribution to the local energy flux of a wave with a higher $l$ is made from
a part of the spectrum at a higher frequency $\omega\sim l^{3/4}$ (solid curves in Fig.~\ref{fig:f3}).
As both of the IGW spectra quickly decline with a growth of $\omega$, our neglect of
waves with the higher spherical degrees is unlikely to lead to a serious mistake.
The same argument justifies our choice of the limited frequency interval
$0.3\,\mu\mbox{Hz}\leq\omega\leq 3\,\mu\mbox{Hz}$ because waves with higher frequencies          
transport negligible amounts of kinetic energy and angular momentum.

Unlike the other papers cited in the preceding section, in this work we have not tried to solve the full problem of
shaping the uniform rotation of the solar radiative core by IGWs. A tentative solution of it has been provided
by \cite{cht05}. We look at this problem from another perspective. Let us assume that the redistribution of
angular momentum in the Sun's core has already established its solid body outer envelope rotation as revealed by the helioseismic
data (e.g., \citealt{cea03}). Despite this, IGWs still continue
to be generated by the envelope convection. We cannot expect {\it a priori} that the stability of
the $\Omega$-profile against perturbations by IGWs strongly depends on its shape. Therefore, we consider it worth investigating
whether these perturbations are sufficiently weak in the present-day Sun for them not to disturb noticeably the Sun's uniform rotation,
provided that the IGWs producing these perturbations had enough energy in the past to couple the Sun's core and envelope rotation.
Given our reformulation of the problem, we have chosen the following initial and boundary conditions for eq. (\ref{eq:ampdf}):
\bea
\label{eq:init}
\Delta\Omega\,(0,r) = \frac{r_{\rm c}-r}{r_{\rm c}-r_{\rm b}}\Delta\Omega\,(0,r_{\rm b})\ \ 
\mbox{for}\ \ r_{\rm b}\leq r\leq r_{\rm c},\\
\mbox{and}\ \ \Delta\Omega\,(t,r_{\rm c}) = 0,\ \ \Delta\Omega\,(t,r_{\rm b})=\Delta\Omega\,(0,r_{\rm b}),
\label{eq:bound}
\eea
where $r_{\rm b}=r_{\rm c}-0.6\,R_\odot$. Thus, we assume that $\Delta\Omega$ initially increases with the depth linearly
up to its maximum value $\Delta\Omega\,(0,r_{\rm b})$. Taking into account that for the present-day Sun $\Omega\,(r_{\rm c})\approx 2.9\,\mu$Hz,
the initial rotational shear in our computations decreases with the depth $z=(r_{\rm c}-r)/R_\odot$ as $q\equiv (\partial\ln\Omega/\partial\ln r)\approx 3.1(x_{\rm c}-z)u_{\rm max}$, where
$x_{\rm c} = r_{\rm c}/R_\odot$, and $u_{\rm max}=\Delta\Omega\,(0,r_{\rm b})/10^{-6}$. In our basic parameter set, we will
use a value of $u_{\rm max} = 10^{-4}\,\mu$Hz\,$\ll\Omega(r_{\rm c})$.

To solve eq. (\ref{eq:ampdf}) one needs to know stellar structure parameters,
such as $\rho$, $N$, $K$, and others, as functions of radius and time. However, because the internal structure of 
the Sun has not changed appreciably in the last billion years, we will use
our model of the present-day Sun as a background for all of our IGW computations
and we will watch that the total integration time in each of them does not exceed $\sim$\,1\,Gyr.
We have solved eq. (\ref{eq:ampdf}) using an original method described in Appendix~A.

\subsection{Constant Viscosity}

Fig.~\ref{fig:f4} shows our results obtained for different values of the constant viscosity $\nu_{0,n}\equiv \nu/10^n$
using the spectra (\ref{eq:zea97}) (panel a) and (\ref{eq:tea02}) (panel b).
This figure illustrates the aforementioned bifurcation
from stationary to oscillatory solutions that occurs at $\nu_{0,8}\approx 5\times 10^{-4}$ in panel a and
at $\nu_{0,8}\approx 8\times 10^{-4}$ in panel b, a sequence of oscillatory solutions (these are analogs of the SLOs
discussed by \citealt{tch05}) for $5\times 10^{-5} \leq \nu_{0,8} \leq 5\times 10^{-4}$ in panel a and
for $7\times 10^{-5} \leq \nu_{0,8} \leq 8\times 10^{-4}$ in panel b, and a transition to
the chaotic behavior of $\Delta\Omega$ at $\nu_{0,8} = 3.8\times 10^{-5}$ in panel a and
at $\nu_{0,8} = 4.9\times 10^{-5}$ in panel b. 
Although in these computations we have only taken into account IGWs with frequencies from
the narrower interval $0.3\,\mu\mbox{Hz}\leq\omega\leq 0.6\,\mu\mbox{Hz}$, this truncation does not
depreciate our results because
waves with these low frequencies carry 75\% and 91\% of the total energy for the spectral distributions
(\ref{eq:zea97}) and (\ref{eq:tea02}), respectively. A comparison of panels a and b shows
that the use of either of the two IGW spectra leads to similar qualitative results.

Since the low-frequency waves get absorbed
very close to the core/envelope interface, we have taken the zooming parameter $k=2.6$ (see Appendix~A) in order to resolve 
the short lengthscale variations of the rotation profile produced by them. For this $k$ and for the used value of $n=8$,
we get a time scaling factor of $9.7\,$yr for our dimensionless PDE (\ref{eq:ampdf2}).
The real timescales of the oscillations of $\Delta\Omega$ are somewhat longer than this because, additionally,
they are inversely proportional to the amplitude of the second term on the right-hand side of the PDE
which is of order $10^{-2}$\,--\,$10^{-3}$. Therefore, the minimum time intervals between consecutive
curves in Fig.~\ref{fig:f4} are $10^3$ yr (panel a) and $10^2$ yr (panel b). 
A decrease of $\nu_0$ leads to both longer timescales and larger amplitudes of the oscillations
because, in order to compete with the disturbing action of IGWs,
a smaller viscosity needs a stronger shear to be built up, which means longer viscous dissipation times.

So far, we have used our basic parameter set that includes a limited number of $l=1,\,2,\,3$, and 4, and uses    
$u_{\rm max} = 10^{-4}\,\mu$Hz in the initial and boundary conditions (\ref{eq:init}\,--\,\ref{eq:bound}).
Panel a in Fig.~\ref{fig:f5} demonstrates that neither the addition of 4 extra $l$ values nor the increase of 
$u_{\rm max}$ by the factor $10^3$, which results in the initial shear $q\approx 0.22$ near
the interface, change much the period and amplitude of the oscillations of $\Delta\Omega$.

Although the changes caused by the increase of $\Delta\Omega(0,r_{\rm b})$ ($u_{\rm max}$)
turn out to be unimportant for our investigation of the ability of IGWs to disturb
the internal uniform rotation of the present-day Sun, they were shown by \cite{tch05} to be a matter 
of great importance in the problem of angular momentum extraction by IGWs from the radiative core of a young
solar-type MS star. Indeed, a steep initial $\Delta\Omega$-profile results in the SLOs that are asymmetric with respect
to the line $\Delta\Omega=0$ (red curves in Fig.~\ref{fig:f5}a). Such SLOs may work as a filter that predominantly
absorbs prograde waves. This means that among low-degree high-frequency waves that arrive at a rapidly rotating central part of the star
retrograde waves transporting negative angular momentum will be over-represented. Hence, when                  
being damped in the core they will spin it down.

\subsection{Shear-Induced Viscosity}

The constant viscosity has been adjusted by hand to get one of the three possible outcomes of the competition
between the disturbing action of IGWs and the smoothing effect of the viscous force on the rotation profile.
It turns out that the quasi-periodic oscillations of $\Delta\Omega$ are settled only when $\nu_0$ takes on a value
from a rather narrow interval. It is unlikely that this accidentally happens in real stars. For the IGW filter
composed of quasi-regular temporal and radial variations of $\Delta\Omega$ near the top of radiative core
to work as proposed by \cite{tch05} some self-regulating mechanism for adjusting the proper viscosity values should
apparently be operating there.  One such mechanism could be a shear-induced viscous friction.
In this case, a shear is readily built up as a result of selective damping of
IGWs. Absorbed waves deposit their angular momentum locally, thus pushing rotation away from its stationary state. 
The viscosity coefficient (\ref{eq:nuv}) used by us is appropriate for describing mixing due to the rotation-induced secular shear
instability (\citealt{mm96}). It develops more easily than the dynamical shear instability but it acts
on a smaller length scale $l_{\rm t}$, such that a turbulent eddy of size $\sim$\,$l_{\rm t}$
can effectively exchange heat with its surroundings while it travels a mean free path of order $\sim$\,$l_{\rm t}$.

\cite{tch05} have implemented this mechanism as follows.
They averaged the turbulent diffusion coefficient (\ref{eq:nuvtz}) over a complete oscillation cycle as well as
over a radial extent of the SLOs using a Gaussian of width $0.2H_P$. Thus obtained stationary viscosity profile
was then used in their IGW computations. As is said in Appendix~B, we actually use the same viscosity (\ref{eq:nuv})
but we allow it to vary with time and we do not average it over radius.
For the same choice of IGW parameters as in the preceding section but
for the shear-induced viscosity (\ref{eq:nuv}) calculated with $f_{\rm v}=1$,
results of our solution of the PDE (\ref{eq:ampdf2}) with the spectrum (\ref{eq:zea97})
are plotted in Fig.~\ref{fig:f5}b. 

It is important to note that, like a few other publications (e.g., \citealt{kmg01}; \citealt{tea02}), the work of \cite{tch05}
only contains a discussion of the SLOs near the core/envelope interface. Unlike them, we have decided
to address the question whether the SLOs die out at greater depth or not. In order to shorten our computation time
(i.e. in order to allow longer time steps)
when solving the PDE (\ref{eq:ampdf2}) in the bulk of radiative core we have bounded the IGW frequency
by the values $0.6\,\mu\mbox{Hz}\leq\omega\leq 3\,\mu$Hz, i.e. we have cut off the IGWs that produce
the SLOs very close to the interface, like those shown in Fig.~\ref{fig:f5}b. In spite of this, our minimum frequency
still approximately equals the lowest frequency $0.5\,\mu$Hz used by \cite{cht05}.
It is important to note that the IGW spectra we use are still normalized by equation
(\ref{eq:neteflux}) over the whole frequency and spherical degree intervals:
$0.3\,\mu\mbox{Hz}\leq\omega\leq 3\,\mu\mbox{Hz}$, and $1\leq l\leq l_{\rm c}=30$.
Because of its rapid decline with an increase of $\omega$ a spectrum normalized with
$\omega_{\rm min} > 0.3\,\mu$Hz would have more energetic high-frequency waves than ours,
hence it would produce even stronger oscillations of $\Delta\Omega$ than those obtained by us.

We have solved equation (\ref{eq:ampdf2}) in the $k$\,$=$\,1-zoomed depth interval $0\leq\hat{z}\leq\hat{z}_{\rm b}=10^k\times(x_{\rm c}-x_{\rm b})$,
where $x_{\rm c}=0.713$ and $x_{\rm b}=x_{\rm c}-0.6$,
for time periods less than 1\,Gyr. Hoping to relate, later on, the angular momentum transport and element mixing by IGWs
in the Sun to canonical extra mixing in low-mass red giants, we have taken the parameter $f_{\rm v}=20$ in
the expression (\ref{eq:nuv}) for the shear-induced viscosity because with about that value \cite{dea06} succeeded in reproducing
evolutionary abundance variations of Li and C in the atmospheres of cluster and field red giants.
We have found that, even after having been enhanced by this large factor, the shear-induced viscosity
fails to extinguish large scale SLOs deeper in the solar radiative core. The blue curve in Fig.~\ref{fig:f6}a represents 
an envelope of oscillation amplitudes of $\Delta\Omega$. The farther inward from the core/envelope interface,
the higher the $\Delta\Omega$ oscillation amplitude
and the longer its characteristic time are.
In the outer half of the radiative core the maximum amplitudes by far exceed the deviations of 
the helioseismic data (red squares with error bars) from the uniform rotation profile. A better agreement with
the experimental data is obtained if we choose $f_{\rm v}=10^3$ (purple curve in panel a).
We have tried this value as well because the 3D hydrodynamic simulations by
\citet{bh01} of turbulent mixing induced by the shear instability have shown
that equation (\ref{eq:nuv0}) may underestimate the coefficient of turbulent
diffusion by three orders of magnitude.  Of course, it is not clear
whether the results of \cite{bh01} can be applied directly to real
stars, given that they were obtained assuming plane-parallel geometry and
without taking into account ``the effects of rotation, nuclear reactions, and
variations in radiative processes''.

However, before making any conclusions from the results of these computations
we have to check if the viscosities induced by the shear flows outlined in panel a
are high enough for the turbulent fluid motions producing them not to be broken down by
the molecular viscosity. For this to be true, the flow Reynolds number $Re=\nu_{\rm v}/\nu_{\rm mol}$
must exceed the critical Reynolds number $Re_{\rm c}\approx 40$ (\citealt{schea00}).
In panel b, we have plotted a time averaged $\nu_{\rm v}$ for the cases of
$f_{\rm v}=20$ (blue curve) and $f_{\rm v}=10^3$ (purple curve). In the same plot,
green curve depicts $\nu_{\rm mol}$ while red curve presents the quantity $Re_{\rm c}\times\nu_{\rm mol}$.
Comparing blue, purple, and red curves in panel b, we conclude that the shear-induced
turbulence can only be sustained near the base of the convection zone
where both the shear $q$ is sufficiently strong thanks to the very short lengthscales of
the SLOs and the quantity proportional to $\nu_{\rm v}f_{\rm v}^{-1}(\Omega q)^{-2}$ 
(eq. \ref{eq:nuv0} and dashed curve in Fig.~\ref{fig:f3})
steeply increases with $r$. This raises the following important question:
what alternative shear dissipating mechanism works in the Sun's outer radiative core
that successfully competes (as follows from the helioseismic data) with the disturbing
action of IGWs?

\subsection{Molecular Viscosity and Ohmic Diffusivity}

Fig.~\ref{fig:f6} shows that the viscosity needed to counteract the distortion of
the solar rotation profile by IGWs should not necessarily be too high.
Taking into account that the blue and purple curves in Fig.~\ref{fig:f6}b
represent the time averaged $\nu_{\rm v}$, whose real values change with time following
the oscillations of $\Delta\Omega$, it seems worth testing if the molecular viscosity
can smooth out the large scale SLOs alone.
The shear-induced turbulent viscosity can only be used down to a depth $z\approx 0.04$\,--\,0.05
because below this region the ratio $\nu_{\rm v}/\nu_{\rm mol}$ becomes smaller
than the critical Reynolds number (Fig.~\ref{fig:f6}b). It is interesting that the size of this region
approximately coincides with the thickness of the solar tachocline in which the latitudinal
differential rotation of the convective envelope is transformed into the quasi-solid body rotation
of the radiative core (e.g., \citealt{sz92,chea98}). 
We want to find out if the viscous force in our computations fails to reduce 
amplitudes of the SLOs in the outer
radiative core to values consistent with the helioseismic data simply because we use
the IGW spectrum (\ref{eq:zea97}) instead of (\ref{eq:tea02}). 

It is possible that this negative result is a function of the assumed IGW spectrum.
In order to respond to these questions,
we have employed the spectrum (\ref{eq:tea02}) and a combined
viscosity $\nu = \nu_{\rm v}\,(f_{\rm v})+f_{\rm mol}\times\nu_{\rm mol}$, where 
$\nu_{\rm v}\,(f_{\rm v})$ is a substitute of equation (\ref{eq:nuv}) in which we set $f_{\rm v}=20$ for
$0\leq z\leq 0.04$ and $f_{\rm v}=0$ for $0.04 < z\leq z_{\rm b}$.
We have also investigated models with more vigorous IGWs.
Furthermore, as \cite{cht05} claimed, the total energy luminosity of IGWs produced by
fluctuating Reynolds stresses in the convective envelope of their solar model was 
$8.5\times 10^{29}$\,erg\,s$^{-1}$. This is about 2.7 times as large as our estimated value
of $L_E(r_{\rm c})\approx 3.2\times 10^{29}$\,erg\,s$^{-1}$. Therefore, we have increased our turbulent Mach number
${\cal M}_{\rm t}=8.3\times 10^{-4}$ by this factor and renormalized
the spectrum (\ref{eq:tea02}) respectively. Results of these computations obtained with
the factor $f_{\rm mol}=2$ are plotted with purple curves in Fig.~\ref{fig:f7}.
For test purposes, we have also repeated these computations with an extended set of the spherical
degree $l=1,2,\ldots,7,8$ (blue curves). Fig.~\ref{fig:f7} shows that
the molecular viscosity (even after it has been doubled) cannot compete with the IGWs
that have been shown by \cite{cht05} to be powerful enough to shape the Sun's solid body rotation.

On the other hand, it turns out that a viscosity proportional to the ohmic diffusivity
$\nu = f_{\rm mag}\times\eta_{\rm mag}$ can extinguish the IGW-induced SLOs everywhere in the solar radiative
core except the tachocline region for a value of $f_{\rm mag}\ga 10$
(Fig.~\ref{fig:f8}). This seemingly pure academic exercise has some sense.
Let us assume that differential rotation in the solar radiative core has been suppressed by magnetic processes,
e.g. like those proposed by \cite{chmg93}, \cite{mlm06}, or \cite{s99} (for more details on the latter, see next section).
However, in order that magnetic fields generated by these processes not to decay too quickly through the ohmic dissipation,
an effective magnetic diffusivity $\eta_{\rm e}$ associated with them must exceed $\eta_{\rm mag}$.

\subsection{Effective Magnetic Viscosity}
\label{sec:effmag}

\cite{s99} has proposed a magnetohydrodynamic mode of angular momentum
transport in radiative zones of
differentially rotating stars. Fluid elements experience large-scale horizontal
displacements caused by an unstable configuration of
the toroidal magnetic field (one consisting of stacks of loops concentric with the rotation axis).
Small-scale vertical displacements of fluid elements are coupled to the horizontal motions,
which can cause both mild mixing and much more effective angular momentum transport.
Spruit's key idea
is that no initial toroidal magnetic field is actually needed   
to drive the instability and mixing  because the unstable field configuration
can be generated and maintained by differential rotation in a process similar to convective dynamo.
The Spruit dynamo cycle consists of two consecutive steps: first,
a poloidal field is generated by the vertical displacements of the unstable toroidal field; second,
the new poloidal field is stretched into a toroidal field by differential rotation.

The Spruit mechanism produces a huge magnetic viscosity $\nu_{\rm e}\ga 10^9$\,cm$^2$\,s$^{-1}$
(\citealt{dp07}) that could indeed prevent
IGWs from disturbing the solar uniform rotation. However, it could produce that uniform rotation itself,
without being assisted by IGWs (\citealt{eea05}).

The original prescription for the effective magnetic diffusivity and viscosity in the model of Spruit's dynamo
has recently been criticized by \cite{dp07} (see also \citealt{zea07}). The principal critical argument is that
\cite{s99} has overestimated the horizontal length scale of the Tayler instability that causes the concentric
magnetic loops to slip sideways. Spruit assumed the length was of order a local stellar radius. \cite{dp07}
suggested that one has to account of the Coriolis force when estimating the instability's horizontal length scale.
This reduces the diffusivity by about three orders of magnitude and results in the following expression:
\bea
\eta_{\rm e}\approx 2\,\frac{K_6\,\Omega_{-6}^2}{(N_T)_{-3}^2}\,q^2,\ \ \mbox{cm}^2\,\mbox{s}^{-1}.
\label{eq:newetae}
\eea
There is no need to do any further computations to understand that, with this revised
diffusivity, magnetic fields generated by the Spruit dynamo in the solar radiative core would immediately be dissipated through
the ohmic diffusivity, hence the whole transport mechanism would not function.
Indeed, equation (\ref{eq:newetae}) gives $\eta_{\rm e}\approx 1.8\times 10^2$\,cm$^2$\,s$^{-1} < \eta_{\rm mag}$
near the base of the solar convection zone even if we take $q\approx 1$ which obviously exceeds the upper limit constrained by the helioseismic data.
The revised prescription may only work to reduce differential rotation in a model of the young Sun in which both $\Omega_{-6}$ and $q$
have much larger values (Denissenkov et al., in preparation).

\section{Conclusion}

Our numerical solutions of the angular momentum transport equation (\ref{eq:ampdf}, or \ref{eq:ampdf2}) have demonstrated that neither the molecular viscosity
nor the shear-induced turbulent viscosity can reduce the large scale oscillations of angular velocity in the solar outer radiative core
caused by selective damping of IGWs, provided that the net energy flux of these waves is
strong enough to shape the Sun's solid body rotation. 
If waves were the sole mechanism, we would therefore expect to see large deviations from rigid rotation.
The amplitudes of these oscillations are found to be too large to agree with the helioseismic data.
Our result holds even when the molecular and shear-induced viscosity are multiplied by large factors.
We have proved that only a viscosity exceeding the ohmic diffusivity by a factor of $\ga$\,10 can smooth out the IGW-induced
oscillations of the rotation profile. This may be an indirect indication 
that some magnetic processes are at work here.
To be more precise, our finding actually satisfies a necessary condition for such processes to work because
the effective magnetic diffusivity $\eta_{\rm e}$ associated with them must exceed the ohmic diffusivity $\eta_{\rm mag}$.
Otherwise magnetic fields generated by them will decay through the ohmic resistivity too quickly.
For example, we have found that 
magnetic torques are strong enough to successfully compete with the action of IGWs
only if the original prescription for the Taylor-Spruit dynamo, in which $\eta_{\rm e}\gg\eta_{\rm mag}$, is used.
However, this particular case turns out to be irrelevant to our problem because the Tayler-Spruit mechanism can
shape the solar solid body rotation alone (\citealt{eea05}), without being assisted by IGWs.

The helioseismic data suggest that either there is an efficient angular momentum transport mechanism
in addition to IGWs that smooths out the SLOs produced by the waves or
the spectral energy distribution of IGWs is different (lower) from those used by us. Although the latter assumption
leads us outside the scope of our formulated problem we will comment on it.
It is possible that strong toroidal magnetic fields in the solar tachocline filter out
the IGWs with the doppler-shifted frequency $\sigma$ above the Alfv\'{e}n frequency (\citealt{kea99,kmg03}).
For the minimum frequency $0.6\,\mu$Hz used in our 
computations of the large scale SLOs in the solar outer radiative core the magnetic field strength
required to prevent the inward wave propagation is about $(3\times 10^5)/l$ Gauss (\citealt{kea99}).
For $l=1$, this corresponds to quite a strong field. If it is present in the tachocline then
the waves with high spherical degrees will be trapped there. However, we have only considered
IGWs with $\omega \geq 0.6\,\mu$Hz and low l values. 
Most of them are likely to propagate below the tachocline. 
In order to find out if an enhanced viscosity
in the tachocline can hinder the propagation of low-degree high-frequency waves into the solar radiative core
we have done test computations in which the shear-induced viscosity was increased by a factor of $10^3$.
Their results plotted with black curves in Fig.~\ref{fig:f7} show that this does not help to solve the problem.

Another, more radical possibility is
that the form of the IGW spectra employed by us is completely wrong. For instance, the IGW spectrum
estimated in the 2D hydrodynamic simulations by \cite{rg05} has a flat energy distribution
which goes three orders of magnitude below the peak luminosity in our Fig.~\ref{fig:f1}.
Apparently, if we applied that spectrum in our computations then even the molecular viscosity
could easily smooth out the SLOs produced by such IGWs. However, it is evident as well that IGWs with
this energy distribution could not produce the uniform rotation of the Sun by its present age
(multiply the ages of the rotation profiles in Fig.~1 from \citealt{cht05} by a thousand).

To summarize, we do not see what microscopic or pure hydrodynamic processes could smooth out the large scale SLOs 
induced by IGWs in the solar outer radiative core.
Therefore, we agree with the conclusion made by \cite{gmci98} about ``the inevitability of 
a magnetic field in the Sun's radiative interior''. 
Indeed, if IGWs are as strong as described by our employed spectra then this magnetic field
is required to trigger magnetic processes that will counteract the disturbing action of IGWs on the
solar rotation profile. On the contrary, if IGWs are weak then we are in need of such magnetic processes
to extract an excess angular momentum from the solar interior on the early MS.

During the preparation of this work, results of a preliminary analysis of the GOLF data on solar g-mode
oscillations have been published by \cite{gea07} suggesting that the solar core at $r\la 0.15\,R_\odot$
may rotate three to five times as fast as the rest of the radiative core. 
If these results prove to be correct, they will seem to rule out
the angular momentum redistribution in the Sun by IGWs, as proposed by \cite{cht05}, because in that case it 
would have been the Sun's inner core to be spun down first. If these results are confirmed, it will mean that the strong $\mu$-gradient
in the Sun's central region has prevented any angular momentum transport mechanism from operating there.
There would also be strong implications for magnetic angular momentum transport, ruling out prescriptions
(such as \citealt{s02}) that predict a weak sensivity to $\mu$ gradients.

\acknowledgements
We acknowledge support from the NASA grant NNG05 GG20G.
PAD thanks Tamara Rogers for usefull discussions and the HAO staff for the warm hospitality.
The National Center for Atmospheric Research is sponsored by the National Science Foundation.

\appendix

\section{Solution of the Angular Momentum Transport Equation}
\label{sec:solution}

In order to shrink the problem's parameter space,
we consider a contribution to the transport of angular momentum only from the waves possessing
the maximum possible azimuthal number $|m|=l$, i.e. from those carrying the maximum angular momentum, both positive
and negative, at given values of $l$ and $\omega$.
To solve the main PDE (\ref{eq:ampdf}), we use a numerical method based on the ideas implemented
by \cite{s90} in his model of the wind flow in the Earth's stratosphere influenced by tropospheric IGWs.

Taking into account the assumed axial symmetry of the IGW spectrum, i.e. that
$S_E(r_{\rm c},\,l,\,m,\,\omega) =  S_E(r_{\rm c},\,l,\,-m,\,\omega)$,
and the fact that the optical depth
in the wave attenuation factor only depends on even powers of the doppler-shifted frequency $\sigma$ (eq. \ref{eq:tau}),
we recast the net angular momentum flux for $m=\pm l$ as
\bea
F_J(r)\equiv \frac{L_J(r)}{4\pi r_{\rm c}^2} = \sum_{l=1}^{l_{\rm max}}\,l\int_{-\omega_{\rm max}}^{\,\omega_{\rm max}}
S_J(l,\omega)\exp\{-\tau(r,l,l,\omega)\}\,d\omega,
\label{eq:netjflux2}
\eea
where
\bea
S_J(l,\omega) \equiv S_E(r_{\rm c},l,l,|\omega|)\,\omega^{-1}.
\label{eq:s}
\eea
In the interval $-\omega_{\rm c} < \omega < \omega_{\rm c}$, we set $S_J(l,\omega)\equiv 0$.
Note that $S_J(l,-\omega) = -S_J(l,\omega)$.

From the computational standpoint, we find it convenient to convert eq. (\ref{eq:ampdf}) to
the following dimensionless form:
\bea
\frac{\partial u}{\partial\hat{t}} = \frac{1}{\rho x^4}
\frac{\partial}{\partial\hat{z}}\left(\rho x^4\nu_n\,\frac{\partial u}{\partial\hat{z}}\right) -
\frac{a_{n,k}}{\rho x^4}\left(\frac{L}{L_\odot}\right)\frac{\partial F}{\partial\hat{z}},
\label{eq:ampdf2}
\eea
where $x = r/R_\odot$, and
\bea
F(\hat{z},u) = \frac{10^3}{\rho_{\rm c}v_{\rm c}^3}\,
\sum_{l=1}^{l_{\rm max}}\,l\int_{-\omega_{\rm max}}^{\,\omega_{\rm max}}S_J(l,\omega)\,
\exp\{-\int_0^{\hat{z}}G(\hat{z}',\,l,\,\omega -lu)\,d\hat{z}'\}\,d\omega.
\label{eq:dfdz}
\eea
In the last equation, $G(\hat{z},\,l,\,\omega -lu) = [l(l+1)]^{3/2}\,\gamma^{-1}(\hat{z})\,(\omega -lu)^{-4}$,
where $\gamma^{-1}(\hat{z}) = 10^{-k}\times 0.2065\times K_6\,N_{-3}\,(N_T)_{-3}^2\,(x_{\rm c} - 10^{-k}\hat{z})^{-3}$
with $x_{\rm c} = r_{\rm c}/R_\odot$. We neglect the square root $\sqrt{1-(\sigma/N)^2}$ in the integrand's
denominator in eq. (\ref{eq:tau}) because we will only consider $\omega_{\rm max}\ll N_{\rm c}$ 
in which case $|\sigma|\equiv |\omega - lu| < 2\omega_{\rm max}\ll N$.

When deriving eqs. (\ref{eq:ampdf2}\,--\,\ref{eq:dfdz}),
we have normalized our basic variables, which are assumed to be initially expressed in cgs units, as follows: 
$u = \Delta\Omega/10^{-6}$, $\nu_n = \nu/10^n$, $\hat{z} = 10^k\times (x_{\rm c}-x)$,
$\hat{t} = (10^{n+2k}/R_\odot^2)\,t$, $K_6 = K/10^6$, and $N_{-3} = N/10^{-3}$. The quantity $k$ is a sort of
zooming parameter. Taking $k > 0$ allows us to look with the scrutiny at results of IGW damping taking place close to the base of
convective envelope, where variations of $u$ may occur on a very short lengthscale of order $10^{-3}$\,--\,$10^{-2}$\,$R_\odot$.
Having done these transformations, it turns out that $a_{n,k} = 1.362\times 10^{8-n-k}$, $\hat{z} = 0$ corresponds
to the core/envelope interface, and $t$ is measured in units of $1.535\times 10^{14-n-2k}$\, yr.
The factor $10^3$ in eq. (\ref{eq:dfdz}) comes about from a combination of 
the factor $0.1$ that estimates the ratio of the convective flux to the total flux from the star
and the factor $10^4$ that represents the reciprocal to the normalization constant for 
the turbulent Mach number that has been included into the coefficient $a_{n,k}$.
Given that $\omega$ is measured in $\mu$Hz, $a_{n,k}$ has additionally been multiplied by the factor $10^6$.
This is necessary to do because the wave frequency that we actually use is $\omega \equiv \omega/10^{-6}$, therefore,
when substituted into eq. (\ref{eq:dfdz}), relation (\ref{eq:s}) should be taken in the form
$S_J = 10^6\times S_E\,(\omega/10^{-6})^{-1}$.

In equations (\ref{eq:ampdf2}\,--\,\ref{eq:dfdz}), all integrals are replaced with series of trapezoids while all derivatives
are approximated by finite differences. The zoomed depth interval $0\leq\hat{z}\leq\hat{z_{\rm b}}$ (here, $\hat{z_{\rm b}}=10^k(r_{\rm c}-r_{\rm b})/R_\odot$)
is divided into $M$ equal subintervals by $M+1$ mesh points while the axisymmetric frequency intervals
$-\omega_{\rm max}\leq\omega\leq-\omega_{\rm c}$ and $\omega_{\rm c}\leq\omega\leq\omega_{\rm max}$, representing
the retrograde and prograde waves, respectively, are divided into $L$ subintervals each.
The resulting system of $M+1$ nonlinear algebraic equations is linearized assuming that after every integration time step $\Delta\,\hat{t}$
the ratio $\max\{|\Delta u_j/u_j|\}_{j=2}^M\ll 1$. In order to speed up the computations, we follow the idea of \cite{s90} to interpolate
functions containing $G(\hat{z_j},l,\omega_i-lu_j)$ in $\hat{z_j}$, $l$, and the combination $(\omega_i-lu_j)$ using
initially prepared and stored tables.\footnote{We take advantage of the fact that $|\omega_i-lu_j|<2\omega_{\rm max}$.}

\section{Viscosity Prescriptions}
\label{sec:visc}

The viscosity $\nu$ in eq. (\ref{eq:ampdf}) plays a very important role because, depending on its value, 
the viscous friction either succeeds or not in smoothing out
oscillations of the rotation profile in the radiative core growing in response to the local deposit of angular momentum that
accompanies the absorption of IGWs. In this work, we employ 5 different prescriptions for $\nu$ as well as some of their
combinations. These are a constant viscosity $\nu_0$,
the molecular viscosity $\nu_{\rm mol}$ (it dominates over the radiative viscosity in the solar-type MS stars),
a viscosity proportional to
the magnetic (ohmic) diffusivity $\eta_{\rm mag}$, a viscosity $\nu_{\rm v}$ associated with vertical turbulence
produced by the secular shear instability induced by differential rotation, and an effective
viscosity $\nu_{\rm e}$ related to magnetic torques generated by the Tayler-Spruit dynamo (\citealt{s99,s02}).

For the viscosity due to the shear-induced vertical turbulence, we use the expression derived by \cite{mm96}
multiplying it by a free parameter $f_{\rm v}$
\bea
\nu_{\rm v} = f_{\rm v}\times\frac{8}{5}\,Ri_{\rm c}\frac{K}{N_T^2}\,\Omega^2q^2,
\label{eq:nuv0}
\eea
where $Ri_{\rm c} = \frac{1}{4}$ is the critical Richardson number, and the shear $q=(\partial\ln\Omega/\partial\ln r)$.
The parameter $f_{\rm v}$ takes into account the fact that, according to hydrodynamic simulations by \cite{bh01},
the original prescription may underestimate the viscosity by the factor of $\sim 10^3$. Alternatively,
\cite{c02} has supposed that $Ri_{\rm c}$ should be at least four times as large as its classical
value.
After the same normalization used in Appendix~A, we have
\bea
(\nu_{\rm v})_n = f_{\rm v}\times\frac{2}{5}\,10^{2k-n}x^2\frac{K_6}{(N_T)_{-3}^2}\left(\frac{\partial u}{\partial\hat{z}}\right)^2.
\label{eq:nuv}
\eea

To compute the effective magnetic viscosity, we use both original Spruit's
equations (see our \sect{sec:effmag}) and the equations revised by \cite{dp07}
\bea
\nu_{\rm e} = \left(\frac{r^2\Omega\eta_{\rm e}^2}{q^2}\right)^{1/3},
\label{eq:nue0}
\eea
where the effective magnetic diffusivity is
\bea
\eta_{\rm e} \approx 2\frac{K}{N_T^2}\,\Omega^2q^2.
\label{eq:etae}
\eea
Combining the last two equations and normalizing the variables, we find
\bea
(\nu_{\rm e})_n = 2.686\times 10^5\times 10^{\frac{2}{3}k-n}\,\Omega_{-6}\left[x^2\frac{K_6}{(N_T)_{-3}^2}\right]^{2/3}
\left(\frac{\partial u}{\partial\hat{z}}\right)^{2/3}.
\label{eq:nue}
\eea

It should be noted that we have implicitly assumed that $N_\mu^2=0$ in eqs. (\ref{eq:nuv0}) and (\ref{eq:etae}).
This approximation may be valid in the outer part of radiative core if we neglect the $\mu$-gradients
produced by the gravitational settling and radiative levitation of chemical elements.
Under this assumption, our choice of $\nu_{\rm v}$ is equivalent to that made by \cite{tch05}.
Indeed, although they have used a prescription proposed by \cite{tz97} 
\bea
\tilde{\nu}_{\rm v} = \frac{8}{5}\,Ri_{\rm c}\frac{K}{N_T^2}\,\Omega^2q^2
\frac{(1+D_{\rm h}/K)}{1+(N_\mu^2/N_T^2)(1+K/D_{\rm h})}
\label{eq:nuvtz}
\eea
that takes into account a reduction of
the stable thermal stratification in the radiative core by strong horizontal turbulence described with
a diffusion coefficient $D_{\rm h}\gg\tilde{\nu}_{\rm v}$, putting $N_\mu^2=0$ in eq. (\ref{eq:nuvtz}) 
and noticing that in evolved solar-type MS stars
$D_{\rm h}\ll K$ (e.g., see Fig.~14 in the paper of \citealt{tch05}) transforms $\tilde{\nu}_{\rm v}$
into our $\nu_{\rm v}$.



\clearpage



\clearpage



\clearpage



\clearpage
\begin{figure}
\epsscale{0.75}
\plotone{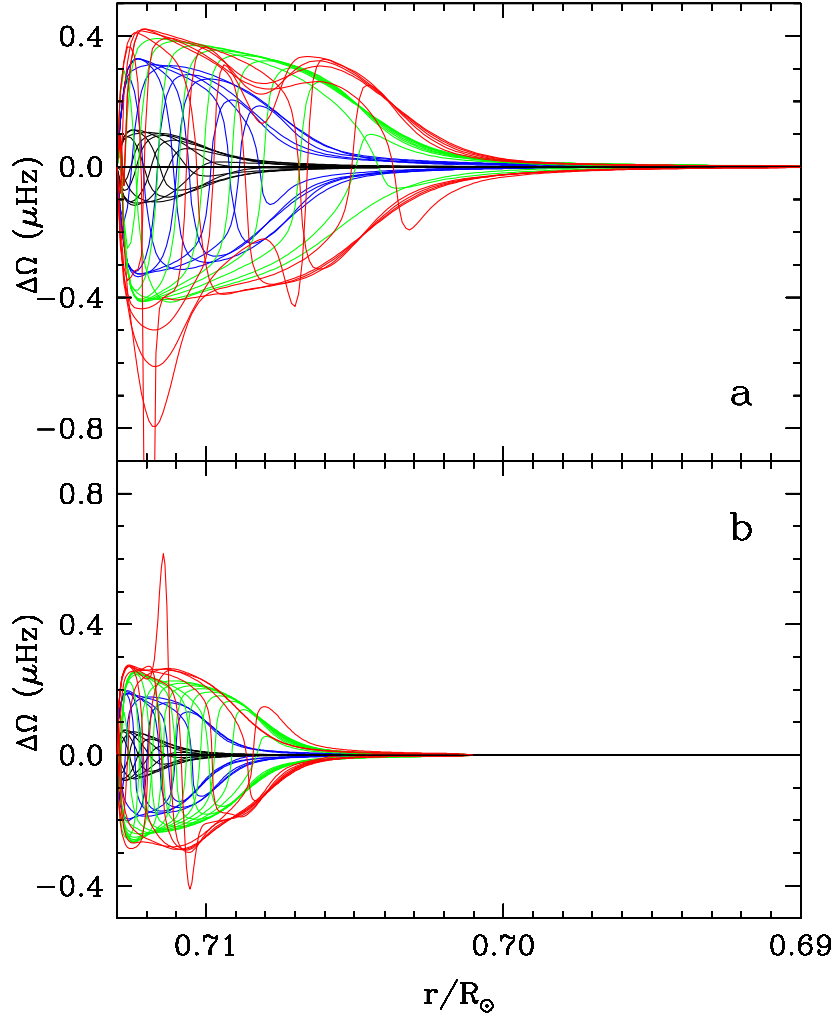}
\caption{The shear-layer oscillations (SLOs) of $\Delta\Omega = \Omega(r)-\Omega_{\rm c}$
         at the top of solar radiative core computed (eq. \ref{eq:ampdf2})
         with the constant viscosity $\nu_{0,8}\equiv \nu_0/10^8$ using
         the IGW spectra (\ref{eq:zea97}) (panel a) and (\ref{eq:tea02})
         (panel b) for the following sets of parameters:
         $0.3\,\mu\mbox{Hz}\leq\omega\leq 0.6\,\mu\mbox{Hz}$;
         $l=1,\,2,\,3,\,4$; $m=\pm l$; and $u_{\rm max}=10^{-4}\,\mu$Hz in the boundary and initial
         conditions (\ref{eq:init}\,--\,\ref{eq:bound}). Panel a:
         black curves --- $\nu_{0,8}=5\times 10^{-4}$, the time interval between
         consecutive curves is $\Delta t=10^3$ yr; 
         blue --- $\nu_{0,8}=10^{-4}$, $\Delta t= 5\times 10^3$ yr;
         green --- $\nu_{0,8}=5\times 10^{-5}$, $\Delta t= 5\times 10^3$ yr;
         red --- $\nu_{0,8}=3.8\times 10^{-5}$, $\Delta t= 10^4$ yr. Panel b:
         black --- $\nu_{0,8}=8\times 10^{-4}$, $\Delta t= 10^2$ yr;
         blue --- $\nu_{0,8}= 10^{-4}$, $\Delta t= 10^3$ yr;
         green --- $\nu_{0,8}= 7\times 10^{-5}$, $\Delta t= 10^3$ yr;
         red --- $\nu_{0,8}= 4.9\times 10^{-5}$, $\Delta t= 2\times 10^3$ yr.
        }
\label{fig:f4}
\end{figure}



\clearpage
\begin{figure}
\epsscale{0.75}
\plotone{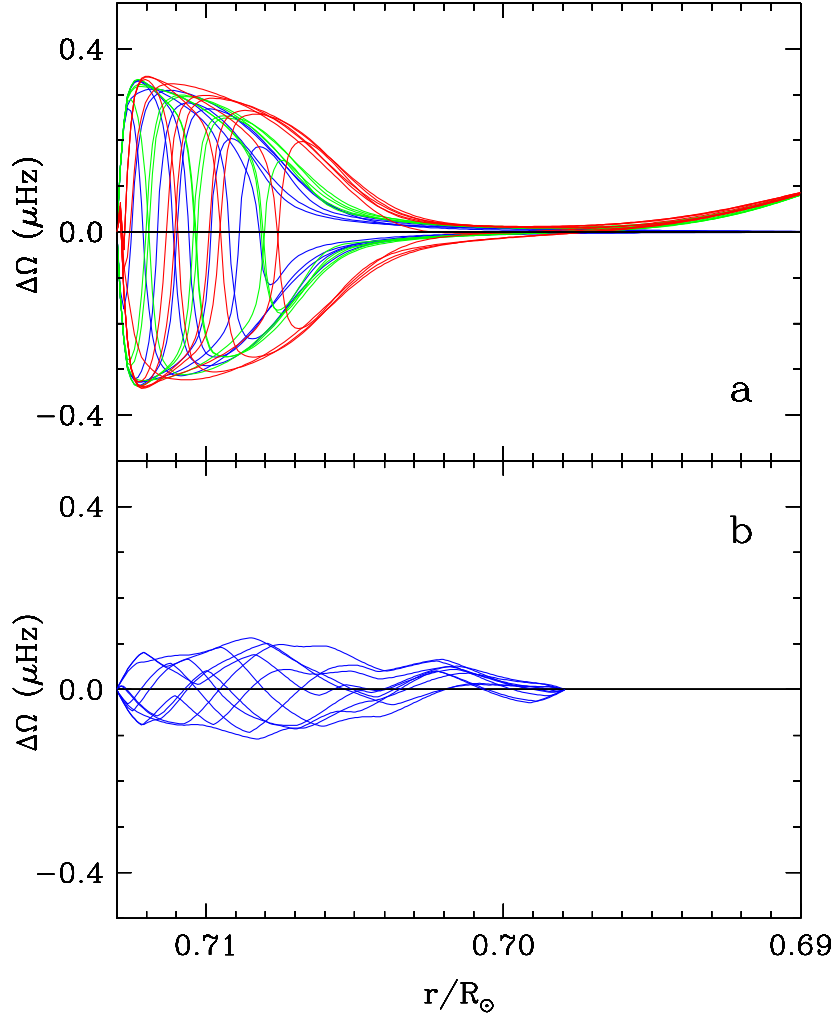}
\caption{The SLOs at the top of solar radiative core computed
         with the constant (panel a) and shear-induced (panel b) 
         viscosity using the IGW spectrum (\ref{eq:zea97})
         and the same sets of basic parameters as in the previous figure.
         Exceptions are the enhanced value of $u_{\rm max} = 0.1\,\mu$Hz
         for green and red curves, and the extended set of
         $l=1,\,2,\ldots,7,\,8$ for red curve. In panel a,
         all curves have
         $\nu_{0,8}= 10^{-4}$, and $\Delta t= 5\times 10^3$ yr.
         Curves in panel b have been computed using the parameter
         $f_{\rm v}= 1$ in eq. (\ref{eq:nuv}), and $\Delta t= 5\times 10^3$ yr.
        }
\label{fig:f5}
\end{figure}



\clearpage
\begin{figure}
\epsscale{0.75}
\plotone{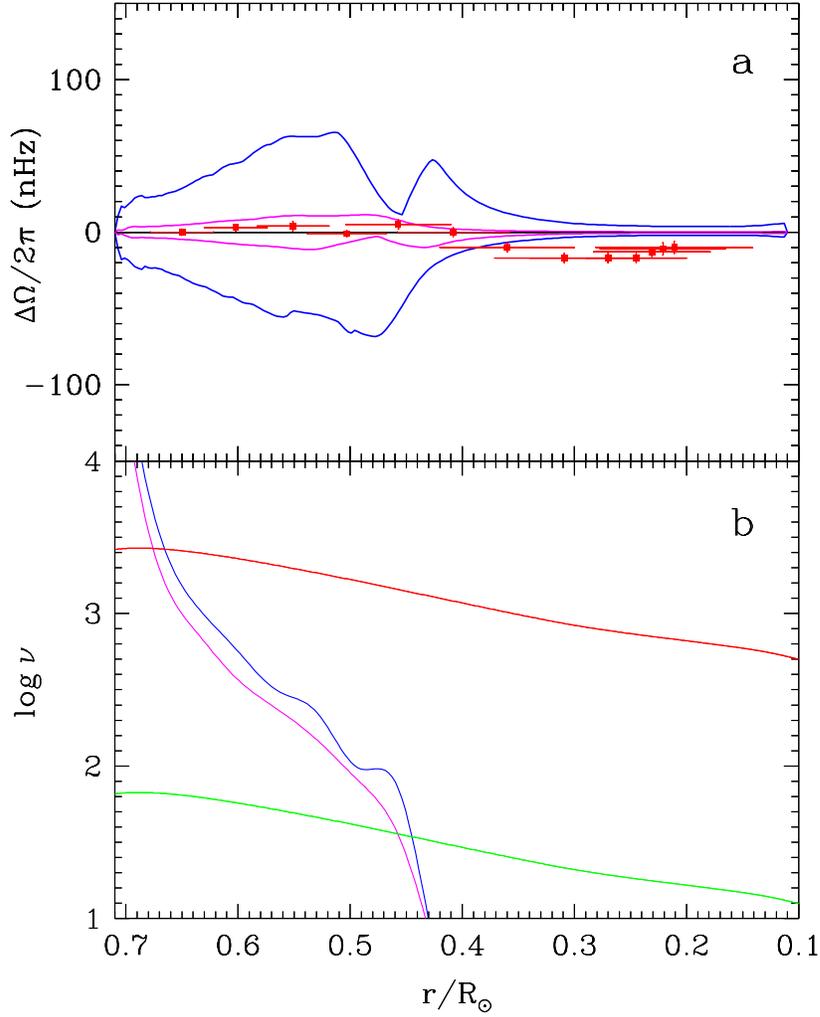}
\caption{Panel a: blue and purple curves are the envelopes of amplitudes of
         the SLOs deep in the solar radiative core computed
         with the shear-induced viscosity (eq. \ref{eq:nuv} with
         $f_{\rm v}=20$ --- blue curve, and with $f_{\rm v}=10^3$ --- purple curve)
         using the IGW spectrum (\ref{eq:zea97}) and
         the same sets of basic parameters as in Fig.~\ref{fig:f4},
         except for the cut-off frequency interval 
         $0.6\,\mu\mbox{Hz}\leq\omega\leq 3\,\mu$Hz.
         Red squares with error bars in panel a are the helioseismic
         data from \cite{cea03}. Panel b: green curve --- the molecular
         viscosity $\nu_{\rm mol}$; red --- the quantity $Re_{\rm c}\times\nu_{\rm mol}$,
         where the critical Reynolds number $Re_{\rm c}=40$ (\citealt{schea00});
         blue and purple curves
         --- the time averaged shear-induced viscosities produced by
         the SLOs whose envelopes are plotted in panel a with the same color.
        }
\label{fig:f6}
\end{figure}



\clearpage
\begin{figure}
\epsscale{0.75}
\plotone{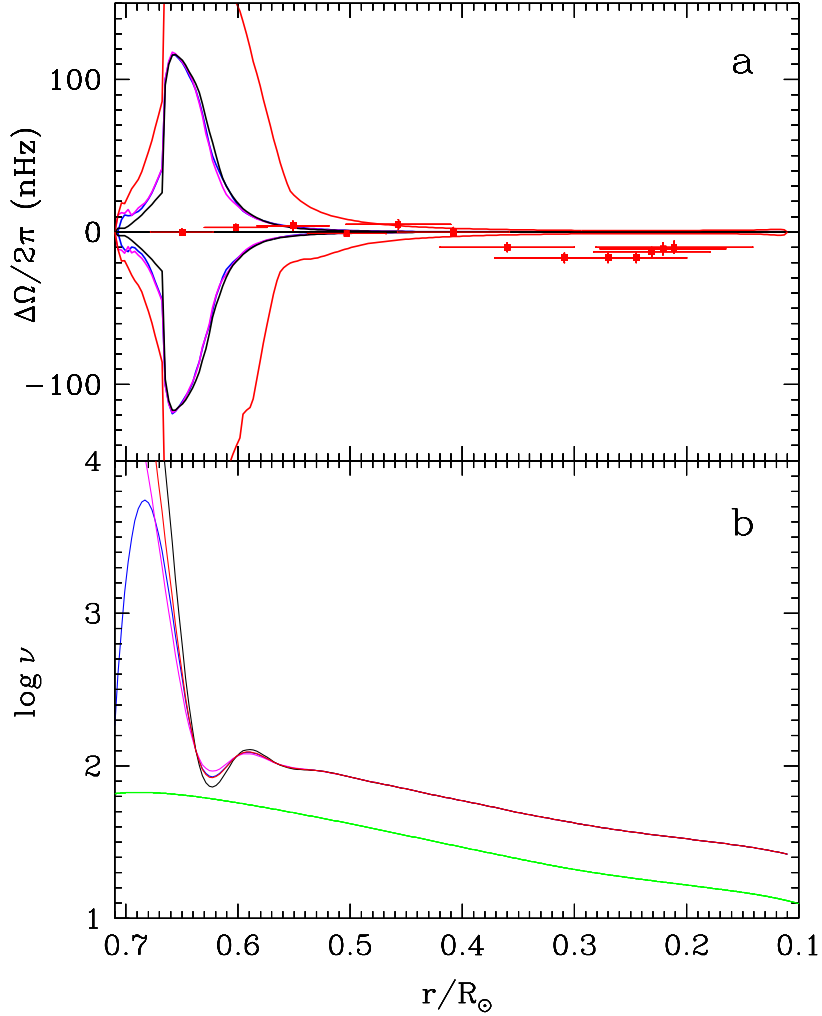}
\caption{Same as in the previous figure but for spectrum (\ref{eq:tea02})
         multiplied by the factor 2.7 (see text)
         and for the combined viscosity $\nu = \nu_{\rm v}\,(f_{\rm v}) + 
         2\nu_{\rm mol}$. Exceptions are blue curves
         that were computed for $l=1,\,2,\ldots,7,\,8$.
         In the expression (\ref{eq:nuv}) for $\nu_{\rm v}$, 
         the following parameters have been used:
         $f_{\rm v}=20$ (blue and purple curves) and 
         $f_{\rm v}=10^3$ (black curve) for $0\leq (r-r_{\rm c})/R_\odot\leq 0.04$,
         while $f_{\rm v}=0$ (all curves) for $(r-r_{\rm c})/R_\odot > 0.04$.
         For comparison, red curves are computed using spectrum (\ref{eq:zea97}).
        }
\label{fig:f7}
\end{figure}



\clearpage
\begin{figure}
\epsscale{0.75}
\plotone{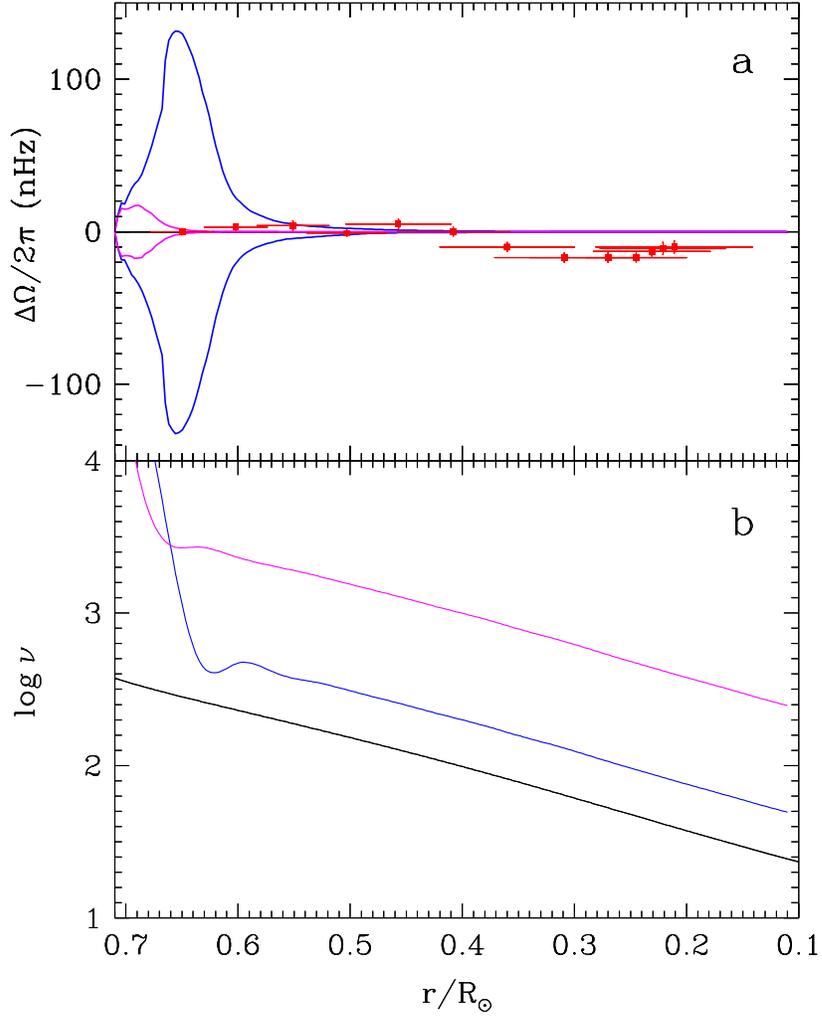}
\caption{Same as in the previous figure but for spectrum (\ref{eq:zea97})
         and for the combined viscosity $\nu = \nu_{\rm v}\,(f_{\rm v}) + 
         f_{\rm mag}\times\eta_{\rm mag}$. In both cases, the parameter
         $f_{\rm v}=20$ has been used in the expression (\ref{eq:nuv}) for $0\leq (r-r_{\rm c})/R_\odot\leq 0.04$. 
         Blue curve is computed with $f_{\rm mag} = 2$, purple --- with
         $f_{\rm mag}=10$. Black curve shows the magnetic diffusivity $\eta_{\rm mag}$.
        }
\label{fig:f8}
\end{figure}



\begin{thebibliography}{}

\bibitem[Barnes(2003)]{b03}
Barnes, S.~A.~2003, ApJ, 586, 464 

\bibitem[Bouvier et al.(1997)]{bea97}
Bouvier, J., Forestini, M., \& Allain, S.~1997, A\&A, 326, 1023

\bibitem[Br\"{u}ggen \& Hillebrandt(2001)]{bh01}
Br\"{u}ggen, M., \& Hillebrandt, W.~2001, MNRAS, 320, 73

\bibitem[Canuto(2002)]{c02}
Canuto, V.~M.~2002, A\&A, 384, 1119

\bibitem[Cattaneo et al.(1991)]{cea91}
Cattaneo, F., Brummell, N.~H., Toomre, J., Malagoli, A., \& Hurlburt, N.~E., 1991, ApJ,
370, 282

\bibitem[Charbonneau \& MacGregor(1993)]{chmg93}
Charbonneau, P., \& MacGregor, K.~B.~1993, ApJ, 417, 762      

\bibitem[Charbonneau et al.(1998)]{chea98}
Charbonneau, P., Tomczyk, S., Schou, J., \& Thompson, M.~J.~1998, ApJ, 496, 1015     

\bibitem[Charbonnel \& Talon(2005)]{cht05}
Charbonnel, C., \& Talon, S.~2005, Science, 309, 2189

\bibitem[Couvidat et al.(2003)]{cea03}
Couvidat, S., Garc\'{\i}a, R.~A., Turck-Chi\`{e}ze, Corbard, T., Henney, C.~J.,
\& Jim\'{e}nez-Reyes, S.~2003, ApJ, 597, L77

\bibitem[Denissenkov \& VandenBerg(2003)]{dvb03}
Denissenkov, P.~A., \& VandenBerg, D.~A.~2003, ApJ, 593, 509

\bibitem[Denissenkov et al.(2006)]{dea06}
Denissenkov, P.~A., Chaboyer, B., \& Li, K.~2006, ApJ, 641, 1087

\bibitem[Denissenkov \& Pinsonneault(2007)]{dp07}
Denissenkov, P.~A., \& Pinsonneault, M.~2007, ApJ, 655, 1157

\bibitem[Eggenberger et al.(2005)]{eea05}
Eggenberger, P., Maeder, A., \& Meynet, G.~2005, A\&A, 440, L9

\bibitem[Garc\'{\i}a L\'{o}pez \& Spruit(1991)]{gls91}
Garc\'{\i}a L\'{o}pez, R.~J., \& Spruit, H.~C.~1991, ApJ, 377, 268

\bibitem[Garc\'{\i}a et al.(2007)]{gea07}
Garc\'{\i}a, R.~A., Turck-Chi\`{e}ze, S., Jim\'{e}nez-Reyes, S.~J., 
Ballot, J., Pall\'{e}, P.~L., Eff-Darwich, A., Mathur, S., \& Provost, J.~2007, Sience, 316, 1591

\bibitem[Goldreich et al.(1994)]{gea94}
Goldreich, P., Murray, N., \& Kumar, P.~1994, ApJ, 424, 466

\bibitem[Gough \& Mcintyre(1998)]{gmci98}
Gough, D., \& Mcintyre, M.~E.~1998, Nature, 394, 755 

\bibitem[Grevesse \& Noels(1993)]{gn93}
Grevesse, N., \& Noels, A.~1993, in Origin and Evolution of the Elements,
ed.. N. Prantzos, E. Vangioni-Flam, \& M. Casse (Cambridge: Cambridge Univ. Press), 15

\bibitem[Kawaler(1988)]{k88}
Kawaler, S.~D.~1988, ApJ, 333, 236

\bibitem[Keppens et al.(1995)]{kea95}
Keppens, R., MacGregor, K.~B., \& Charbonneau, P.~1995, A\&A, 294, 469

\bibitem[Kim \& MacGregor(2001)]{kmg01}
Kim, E.-J., \& MacGregor, K.~B.~2001, ApJ, 556, L117

\bibitem[Kim \& MacGregor(2003)]{kmg03}
Kim, E.-J., \& MacGregor, K.~B.~2003, ApJ, 588, 645 

\bibitem[Kiraga et al.(2005)]{kea05}
Kiraga, M., Stepien, K., \& Jahn, K.~2005, AcA, 55, 205

\bibitem[Krishnamurthi et al.(1997)]{kea97}
Krishamurthi, A., Pinsonneault, M.~H., Barnes, S., \& Sofia, S.~1997, ApJ, 480, 303

\bibitem[Kumar \& Quataert(1997)]{kq97}
Kumar, P., \& Quataert, E.~J.~1997, ApJ, 475, L143

\bibitem[Kumar et al.(1999)]{kea99}
Kumar, P., Talon, S., \& Zahn, J.-P.~1999, ApJ, 520, 859

\bibitem[Lighthill(1978)]{l78}
Lighthill, J.~1978, Waves in Fluids, (Cambridge: Cambridge Univ. Press), Chap. 5

\bibitem[Lindzen \& Holton(1968)]{lh68}
Lindzen, R.~S., \& Holton, J.~R.~1968, J. Atmos. Sci., 25, 1095

\bibitem[Maeder \& Meynet(1996)]{mm96}
Maeder, A., \& Meynet, G.~1996, A\&A, 313, 140

\bibitem[Matt \& Pudritz(2008)]{mp08}
Matt, S., \& Pudritz, R.~E.~2008, arXiv:0801.0436v2 [astro-ph]

\bibitem[Menou \& Le Mer(2006)]{mlm06}
Menou, K., \& Le Mer, J.~2006, ApJ, 650, 1208

\bibitem[Montalb\'{a}n \& Schatzman(2000)]{ms00}
Montalb\'{a}n, J., \& Schatzman, E.~2000, A\&A, 354, 943

\bibitem[Plumb(1977)]{p77}
Plumb, R.~A.~1977, J. Atmos. Sci., 34, 1847

\bibitem[Press(1981)]{p81}
Press, W.~H.~1981, ApJ, 245, 286

\bibitem[Rieutord \& Zahn(1995)]{rz95}
Rieutord, M., \& Zahn, J.-P.~1995, A\&A, 296, 127 

\bibitem[Ringot(1998)]{r98}
Ringot, O.~1998, A\&A, 335, L89

\bibitem[Rogers \& Glatzmaier(2005)]{rg05}
Rogers, T.~M., \& Glatzmaier, G.~A.~2005, MNRAS, 364, 1135

\bibitem[Rogers \& Glatzmaier(2006)]{rg06}
Rogers, T.~M., \& Glatzmaier, G.~A.~2006, ApJ, 653, 756 

\bibitem[Saravanan(1990)]{s90}
Saravanan, R.~1990, J. Atmos. Sci., 47, 2465

\bibitem[Schatzman et al.(2000)]{schea00}
Schatzman, E., Zahn, J.-P., \& Morel, P.~2000, A\&A, 364, 876

\bibitem[Sestito \& Randich(2005)]{sr05}
Sestito, P., \& Randich, S.~2005, A\&A, 442, 615

\bibitem[Sills \& Pinsonneault(2000)]{sp00}
Sills, A., \& Pinsonneault, M.~H.~2000, ApJ, 540, 489

\bibitem[Spiegel \& Zahn(1992)]{sz92}
Spiegel, E.~A., \& Zahn, J.-P.~1992, A\&A, 265, 106

\bibitem[Spruit(1999)]{s99}
Spruit, H.~C.~1999, A\&A, 349, 189

\bibitem[Spruit(2002)]{s02}
Spruit, H.~C.~2002, A\&A, 381, 923

\bibitem[Stauffer \& Hartmann(1987)]{sh87}
Stauffer, J.~R., \& Hartmann, L.~W.~1987, ApJ, 318, 337

\bibitem[Stein \& Nordlund(1989)]{sn89}
Stein, R.~F., \& Nordlund, A.~1989, ApJ, 342, L95

\bibitem[Talon \& Zahn(1997)]{tz97}
Talon, S., \& Zahn, J.-P.~1997, A\&A, 317, 749 

\bibitem[Talon et al.(2002)]{tea02}
Talon, S., Kumar, P., \& Zahn, J.-P.~2002, ApJ, 574, L175

\bibitem[Talon \& Charbonnel(2003)]{tch03}
Talon, S., \& Charbonnel, C.~2003, A\&A, 405, 1025

\bibitem[Talon \& Charbonnel(2005)]{tch05}
Talon, S., \& Charbonnel, C.~2005, A\&A, 440, 981

\bibitem[Yoden \& Holton(1988)]{yh88}
Yoden, S., \& Holton, J.~R.~1988, J. Atmos. Sci., 45, 2703

\bibitem[Zahn et al.(1997)]{zea97}
Zahn, J.-P., Talon, S., \& Matias, J.~1997, A\&A, 322, 320

\bibitem[Zahn et al.(2007)]{zea07}
Zahn, J.-P., Brun, A.~S., \& Mathis, S.~2007, A\&A, 474, 145

\end{thebibliography}
\end{document}